\documentclass{emulateapj}
\usepackage{color} 

\usepackage{epsfig}
\usepackage{subfigure}
\usepackage{graphicx}
\usepackage{amsmath}
\usepackage{natbib}
\usepackage{amssymb}
\usepackage{amsmath}
\usepackage{mathrsfs}

\newcommand{\mb}{\boldsymbol}

\shorttitle{Hall-Controlled Gas Dynamics in PPDs}
\shortauthors{X. N. Bai}

\begin{document}


\title{Hall-effect Controlled Gas Dynamics in Protoplanetary Disks ---
II: Full 3D Simulations toward the Outer Disk}


\author{Xue-Ning Bai\altaffilmark{1}}

\affil{Institute for Theory and Computation, Harvard-Smithsonian
Center for Astrophysics, 60 Garden St., MS-51, Cambridge, MA 02138}
\email{xbai@cfa.harvard.edu}

\altaffiltext{1}{Hubble Fellow}




\begin{abstract}
We perform 3D stratified shearing-box MHD simulations on the gas dynamics of
protoplanetary disks threaded by net vertical magnetic field $B_{z0}$. All three non-ideal
MHD effects, Ohmic resistivity, the Hall effect and ambipolar diffusion are included in a
self-consistent manner based on equilibrium chemistry. We focus on regions toward
outer disk radii, from 5-60 AU, where Ohmic resistivity tends to become negligible,
ambipolar diffusion dominates over an extended region across disk height, and the
Hall effect largely controls the dynamics near the disk midplane. We find that around
$R=5$ AU, the system launches a laminar/weakly turbulent magnetocentrifugal wind
when the net vertical field $B_{z0}$ is not too weak, as expected. Moreover, the wind
is able to achieve and maintain a configuration with reflection symmetry at disk
midplane, as adopted in our previous work. The case with anti-aligned field polarity
(${\mb\Omega}\cdot{\mb B}_{z0}<0$) is more susceptible to the MRI when $B_{z0}$
drops, leading to an outflow oscillating in radial directions and very inefficient angular
momentum transport.
At the outer disk around and beyond $R=30$ AU, the system shows vigorous MRI
turbulence in the surface layer due to far-UV ionization, which efficiently drives disk
accretion. The Hall effect affects the stability of the midplane region to the MRI, leading
to strong/weak Maxwell stress for aligned/anti-aligned field polarities. Nevertheless,
the midplane region is only very weakly turbulent, where the vertical rms velocity is on
the order of $10^{-2}$ sound speed. Overall, the basic picture is analogous to the
conventional layered accretion scenario applied to the outer disk. In addition, we
find that the vertical magnetic flux is strongly concentrated into thin, azimuthally
extended shells in most of our simulations beyond $\sim15$ AU when $B_{z0}$ is
not too weak. This is a generic phenomenon {\it unrelated} to the Hall effect, and
leads to enhanced zonal flow. Future global simulations are essential in determining
the outcome of the disk outflow, magnetic flux transport, and eventually the global
disk evolution.
\end{abstract}



\section{Introduction}\label{sec:intro}

The gas dynamics in protoplanetary disks (PPDs) is largely controlled by non-ideal
magnetohydrodynamics (MHD) effects due to the weak level of ionization, which 
include Ohmic resistivity, the Hall effect and ambipolar diffusion (AD). The three effects
co-exist in PPDs, with Ohmic resistivity dominating dense regions (midplane region
of the inner disk), AD dominating tenuous regions (inner disk surface and the outer
disk), and the Hall dominated region lies in between. While Ohmic resistivity and AD
have been studied extensively in the literature, the role of the Hall effect remains
poorly understood. This paper is the continuation of our exploration on the role of the
Hall effect in PPDs, following Bai (2014, hereafter, paper I), where extensive
summary of the literature and background information were provided in great detail.

One of the major new elements introduced by the Hall effect is that the gas dynamics
depends on the polarity of the external poloidal magnetic field (${\mb B}_0$) threading
the disk. Such external field is expected to be present in PPDs as inherited from the star
formation process (see \citealp{MckeeOstriker07} and \citealp{Crutcher12} for an
extensive review), and is also required to explain the observed accretion rate in PPDs
\citep{BaiStone13b,Bai13,Simon_etal13a,Simon_etal13b}. Observationally, the
large-scale magnetic fields have been found to thread star-forming cores
\citep{Chapman_etal13,Hull_etal14}, and it is conceivable that the large-scale field with
${\mb B}_0\cdot{\mb\Omega}>0$ and ${\mb B}_0\cdot{\mb\Omega}<0$ are equally possible,
where ${\mb\Omega}$ is along the disk rotation axis. At the scale of PPDs, particularly
the scale where the Hall term is dynamically important ($\lesssim50-60$ AU), one would
expect different physical consequences for different field polarities.

In paper I, we focused on the inner region of PPDs ($R\lesssim15$ AU), where the
midplane region is dominated by Ohmic resistivity and the Hall effect, and the disk upper
layer is dominated by AD. Without including the Hall effect, it has been found that the
magnetorotational instability (MRI, \citealp{BH91}) is completely suppressed in the inner
disk, leading to a laminar flow and the disk launches a magnetocentrifugal wind
(\citealp{BaiStone13b}, \citealp{Bai13}). With the inclusion of the Hall effect studied in
paper I, the basic picture of laminar wind still holds, but the radial range where a laminar
wind solution can be found depends on the magnetic polarity: for ${\mb B}_0\cdot{\mb\Omega}>0$,
range of stable wind solution is expected to extend to $R\sim10-15$ AU, while for
${\mb B}_0\cdot{\mb\Omega}<0$, the stable region is reduced to only up to $\sim3-5$ AU.
In addition, horizontal magnetic field is amplified/suppressed in the two cases as a result
of the interplay between the Hall effect and shear (see also \citealp{Kunz08,Lesur_etal14}).

The studies in paper I predominantly use quasi-1D simulations to construct the laminar
wind solutions. In this paper, we shift toward the outer PPDs and consider regions beyond
which the MRI is expected to set in (3-15 AU depending on the strength and polarity
of ${\mb B}_0$), up to the radius where the Hall effect has significant influence ($\sim60$
AU), and conduct full 3D simulations to accommodate turbulent fluctuations and potentially
large-scale variations. In this range of disk radii, the
midplane region is largely dominated by both the Hall effect and AD, and AD becomes
progressively more dominated toward disk surface layer. Without including the Hall effect,
it was found that the MRI is able to operate in the AD dominated midplane though the level
of turbulence is strongly reduced due to AD \citep{Bai13,Simon_etal13b}. In addition, as
the far-UV (FUV) ionization penetrates deeper (geometrically) into the disk, MRI operates
much more efficiently in the much-better-ionized surface FUV layer
\citep{PerezBeckerChiang11b,Simon_etal13b}, which carries most of the accretion flow. The
inclusion of the Hall effect is expected to modify the gas dynamics in the disk midplane region,
which should also be controlled by the polarity of the large-scale magnetic field.

We begin by studying the properties of the MRI in the presence of both the Hall effect and
AD using unstratified shearing-box simulations and discuss its relevance in PPDs in Section
2. In Sections 3 we describe the numerical set up for our full 3D stratified simulations of PPDs
with realistic ionization profiles and run parameters. In Sections 4 and 5, we present simulation
results at two focused radii, 30 AU (Section 4) and 5 AU (Section 5), emphasizing the role
played by the Hall effect. We briefly discuss simulations at other disk radii (15 and 60 AU) in
Section 6 which help map out the dependence of PPD gas dynamics on disk radii. We
summarize the main results and discuss observational consequences, caveats and future
directions in Section 7.

\section[]{MRI with Hall Effect and Ambipolar Diffusion}

In this section, we focus on the general properties on the non-linear evolution of the MRI
in the presence of both the Hall effect and AD, applicable to the outer region of PPDs,
which serve to guide more realistic simulations for the rest of this paper. All simulations
are performed using the ATHENA MHD code \citep{Stone_etal08}, with the relevant
non-ideal MHD terms implemented in our earlier works (\citealp{BaiStone11}, paper I).
We adopt the shearing-sheet framework \citep{GoldreichLyndenbell65} without
including vertical gravity (hence vertically unstratified). Here, dynamical equations are
written in Cartesian coordinate in the corotating frame with a local disk patch with angular
velocity $\Omega{\mb e}_z$. As a convention, ($x, y, z$) represent radial, azimuthal
and vertical coordinates respectively. The equations are the same as Equations (2)-(5)
in paper I, except the $\Omega^2z{\mb e}_z$ term in the momentum equation, and the
Ohmic resistivity term in the induction equation are dropped. An isothermal equation of
state $P=\rho c_s^2$ is adopted with $c_s$ being the sound speed. In code unit, we
have $\rho_0=c_s=\Omega=1$, where $\rho_0$ is the initial gas density (or midplane
density for stratified simulations in the following sections). The unit for magnetic field is
chosen such that magnetic permeability $\mu=1$.

In the following, we first discuss the relative importance of the Hall effect and AD
in the relevant regions of PPDs. We then discuss the MRI linear dispersion relation of
in the presence of both the Hall and AD terms. Finally, we proceed to non-linear unstratified
shearing-box simulations. Our survey of the parameter space is by no means complete,
but we have chosen the range of parameters that are most relevant to the regions of
PPDs that we study in the Sections that follow (midplane regions up to $\sim60$ AU).

\subsection[]{Relative Importance of the Hall Effect and Ambipolar Diffusion in PPDs}

The Hall effect is characterized by a physical scale, and in the absence of charged
grains, it reads \citep{KunzLesur13}
\begin{equation}
l_H\equiv\frac{v_A}{\omega_H}=\bigg(\frac{\rho}{\rho_i}\bigg)
\bigg(\frac{v_A}{\omega_i}\bigg)\ ,
\end{equation}
where $v_A=B/\sqrt{4\pi\rho}$ is the Alfv\'en velocity, $\omega_i$ is the ion cyclotron
frequency, $\omega_H=(\rho_i/\rho)\omega_i$ is the Hall frequency, $\rho_i$ and
$\rho$ are the mass densities of the ions and the bulk of the gas, respectively,
with $\rho_i\ll\rho$ for weakly ionized gas. Note that both $v_A$ and $\omega_i$
are proportional to the magnetic field strength, hence $l_H$ is field-strength
independent, and is determined solely by the ionization fraction. In the disks, it
is natural to normalize $l_H$ by the disk scale height $H\equiv c_s/\Omega$.
The associated Hall diffusivity $\eta_H$ can be expressed as
\begin{equation}
\eta_H=v_Al_H\ .
\end{equation}
Note that $\eta_H\propto B$.

Ambipolar diffusion is characterized by the frequency that neutrals collide with
ions $\gamma_i\rho_i$, where $\gamma_i$ is the coefficient of momentum transfer
for ion-neutral collisions. In the disk, it is natural to normalize $\gamma_i\rho_i$ to
the disk orbital frequency, by defining
\begin{equation}
Am\equiv\frac{\gamma_i\rho_i}{\Omega}\ ,\label{eq:Am}
\end{equation}
which is the Elssaser number for AD. Generally, AD plays an important role
in the gas dynamics when $Am\lesssim10$ \citep{BaiStone11}. The associated
AD diffusivity is given by
\begin{equation}
\eta_A=v_A^2/\gamma_i\rho_i\ .
\end{equation}
Note that $\eta_A\propto B^2$.

The above definitions apply when electrons and ions are the main charged
species, where the physics can be described most easily. Generalizations to
include charged grains can be found in, e.g., \citet{Wardle07} and \citet{Bai11a}
which are used in our vertically stratified simulations in subsequent sections.

Jointly, we see that the product of the two dimensionless numbers $l_H/H$
and $Am$ is independent of the ionization fraction, and is given by
\begin{equation}
Am\cdot\bigg(\frac{l_H}{H}\bigg)
=\frac{\gamma_i\rho}{\omega_i}\bigg(\frac{v_A}{c_s}\bigg)\ .
\propto\frac{\sqrt{\rho}}{c_s}
\end{equation}
When adopting the minimum-mass solar nebula disk model (MMSN,
\citealp{Weidenschilling77,Hayashi81}), we have that at the disk midplane,
$\rho_0\propto R^{-11/4}$, $c_s\propto R^{-1/4}$, hence
$Am\cdot(l_H/H)\propto R^{-9/8}$. More specifically, we find\footnote{The
factor $\gamma_i$ and $\omega_i$ depend on the mass of the ions. However,
for the ion mass $m_i\gg m_H$, the dependence diminishes. The value
computed here assumes the gas mean molecular weight $\mu=2.33m_H$,
following the formulas in \citet{Bai11a}.} 
\begin{equation}\label{eq:AmlH}
Am\cdot\bigg(\frac{l_H}{H}\bigg)\approx0.64\ \bigg(\frac{R}{10{\rm AU}}\bigg)^{-9/8}\ .
\end{equation}
In the outer region of PPDs, the value of $Am$ is found to be of order unity for
a wide range of disk radii \citep{Bai11a,Bai11b}, and this formula provides a very
useful relation in estimating the importance of AD and the Hall effect in PPDs. If we
consider the Hall effect to be important when $l_H/H\gtrsim0.1$, then the influence
of the Hall effect extends to $\sim50-60$ AU.

For the MRI, the relative importance of the Hall effect and AD is characterized by
their respective Elsasser numbers, defined as $v_A^2/\eta\Omega$, with $\eta$
being the respective diffusivities for the Hall effect ($\eta_H$) and AD ($\eta_A$). With
the AD Elssaser number introduced in (\ref{eq:Am}), the Hall Elsasser number can
be written as (see paper I for details)
\begin{equation}
\chi\equiv\frac{\omega_H}{\Omega}\ .
\end{equation}
Note that $\chi$ depends on field strength ($\propto B$), and also
\begin{equation}
\frac{l_H}{H}=\frac{1}{\chi}\frac{v_A}{c_s}=\frac{1}{\chi}\sqrt{\frac{2}{\beta}}
=\frac{X}{\sqrt{2\beta}}\ ,\label{eq:convert}
\end{equation}
where the plasma $\beta=8\pi P/B^2$ is the ratio of gas to magnetic pressure,
and $X\equiv2/\chi$ is another commonly adopted quantity in the literature
\citep{SanoStone02a,SanoStone02b}. The comparison between $\chi$ and
$Am$ reveal the relative importance between the Hall effect and AD, and the Hall
term becomes comparably less dominant for larger $\chi$ (stronger magnetic field
and smaller density). Using Equation (\ref{eq:AmlH}), we find
\begin{equation}\label{eq:Amchi}
\frac{Am}{\chi}\approx4.5\sqrt{\frac{\beta}{100}}\bigg(\frac{R}{10{\rm AU}}\bigg)^{-9/8}\ .
\end{equation}
Again, we see that $Am$ and $\chi$ are likely of the same order for a wide range
of disk radii given the typical magnetic field strength of $\beta\lesssim100$
(saturated $\beta$) in the outer disk.

In our definition, $\omega_H$, $l_H$ and $\chi$ are all positive. On the other hand,
the Hall effect also depends on the polarity of the magnetic field relative to
${\mb\Omega}$. To distinguish the two cases, we always state explicitly the polarity
of the background magnetic field $B_{z0}>0$ or $B_{z0}<0$ for fields aligned and
anti-aligned with ${\mb\Omega}$ in this paper.

\begin{figure*}
    \centering
    \includegraphics[width=180mm]{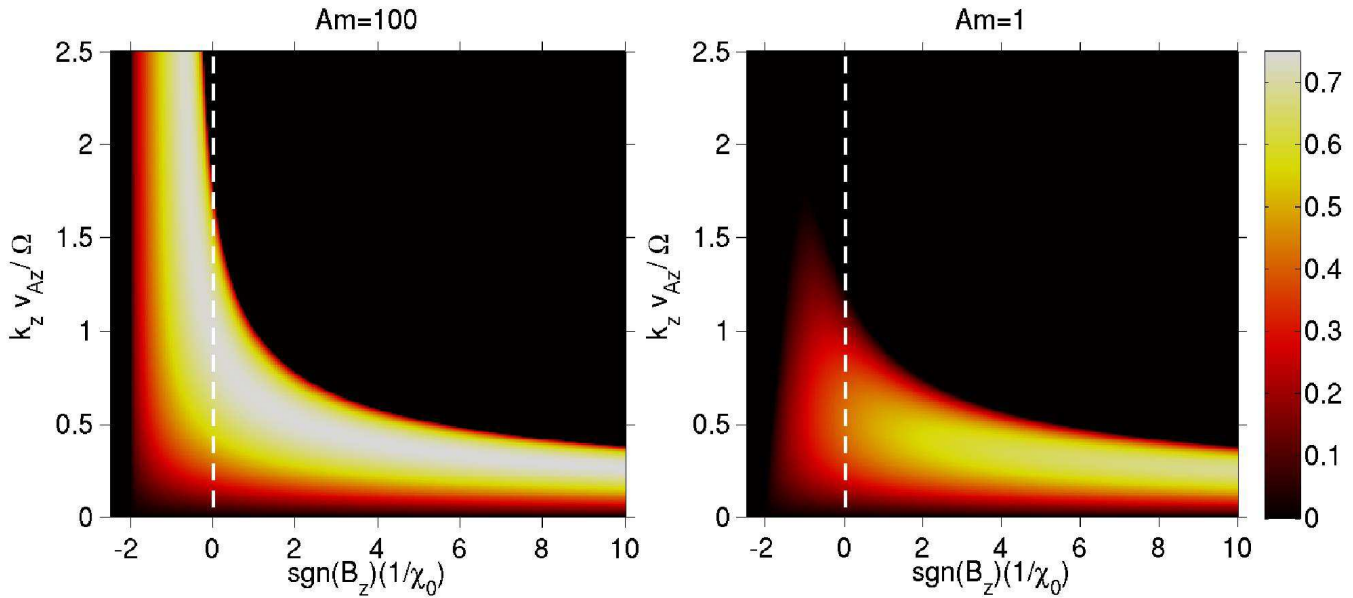}
  \caption{Linear growth rate of the MRI in the presence of the Hall effect and AD, in
  the case of pure vertical background magnetic field, and for modes with pure
  vertical wavenumbers $k_z=k$. Growth rate is drawn as a function of normalized
  wavenumber $kv_A/\Omega$ and sgn$(B_z)(1/\chi_0)$, with two panels showing
  results for fixed $Am=100$ (ideal MHD) and $Am=1$ (strong AD). Note that no
  unstable mode exists for sgn$(B_z)(1/\chi_0)\leq-2$. }\label{fig:MRIdisp}
\end{figure*}

\subsection[]{Linear Properties}

The linear dispersion relation of the MRI for general axisymmetric perturbations
in the Hall and AD regimes has been derived separately in \citet{BalbusTerquem01}
and \citet{KunzBalbus04,Desch04}. The authors considered a general background
field configuration ${\mb B}_0=B_{z0}{\mb e}_z+B_{\phi0}{\mb e}_{\phi}$, and general
axisymmetric perturbations of the form $\exp{({\rm i}{\mb k}\cdot{\mb x}+\sigma t)}$
with ${\mb k}=k_x{\mb e}_x+k_z{\mb e}_z$. The main results reveal that for the MRI
modes, the Hall term is coupled only to the vertical magnetic field, while the AD term
is also coupled to the toroidal magnetic field. As a result, the presence of a background
toroidal field has little effect on the Hall MRI, but facilitates the MRI to operate in the AD
dominated regime with $Am\lesssim1$.
A joint dispersion relation including all non-ideal MHD terms was given by
\citet{PandeyWardle12}. It was shown that while contributions from the Hall and AD
terms are independent, the joint effect is that regimes stable to pure Hall-MRI can be
rendered unstable due to AD, a situation which again requires net toroidal field
and strong AD ($Am\lesssim1$).

Exploring the full parameter space of the MRI in the presence of Hall and AD effects with
different field orientations with non-linear simulations is beyond the scope of this work.
Here, we restrict ourselves to pure vertical background field with either $B_{z0}>0$ or $B_{z0}<0$.
This choice makes the dispersion relation much simpler, where the most unstable mode has
pure vertical wavenumber $k_z=k$, and for these modes, AD behaves the same way as Ohmic
resistivity by replacing $\eta_A$ with $\eta_O$, in the linear regime. This case also covers the
most essential MRI physics relevant to PPDs, since the Hall term is not directly coupled to the
toroidal field, and for AD, the background toroidal field does not strongly affect the level of the
MRI turbulence for $Am\gtrsim1$ \citep{BaiStone11}.

In reference to previous works (e.g., \citealp{Wardle99}), we show in Figure \ref{fig:MRIdisp}
the MRI growth rate for pure vertical modes $k=k_z$ as a function of dimensionless
wavenumber $kv_{A0}/\Omega$ and $1/\chi_0$, where subscript `$_0$' represents $\chi$
and $v_A$ determined from background field, and similarly we use $\beta_0$ to denote
plasma $\beta$ for the background field. Magnetic polarity is reflected using sgn$(B_{z0})$.
We consider two cases with $Am=1$ and $Am=100$.

For $Am=100$ (very weak AD), the dispersion relation is well described by pure Hall
MRI. For $B_{z0}>0$, the most unstable mode always has the maximum
growth rate of $0.75\Omega^{-1}$, and the most unstable wavelength $\lambda_m$
shifts progressively to larger scales with $\lambda_m\propto\chi_0^{-1/2}$ as the Hall
term strengthens ($\chi_0\rightarrow0$). Normalizing
to the disk scale height, we find
\begin{equation}
\frac{\lambda_m}{H}\approx4\pi\sqrt{\frac{l_H}{3H}}\bigg(\frac{2}{\beta_0}\bigg)^{1/4}
\approx0.5\sqrt{\frac{3l_H}{H}}\bigg(\frac{10^4}{\beta_0}\bigg)^{1/4}\ .\label{eq:lambdam}
\end{equation}
For $B_{z0}<0$, unstable modes exist only when $(1/\chi_0)<2$, and
unstable wavenumber can extend virtually to infinity when $(1/\chi_0)>1/2$.

For $Am=1$, we see that small-scale modes are strongly suppressed. For $B_{z0}<0$,
the most unstable modes have wave numbers of $kv_A/\Omega\sim0.5$.  In the absence
of the Hall effect ($1/\chi_0=0$), $\lambda_m$ is increased by a factor of $\sim2$ due AD.
For $B_{z0}>0$ and toward stronger Hall term ($1/\chi_0\gtrsim5$), $\lambda_m$ is less
affected by AD since it is shifted to larger scales, and the maximum growth rate is only
slightly reduced.

\subsection[]{Unstratified Shearing-box Simulations}

Our unstratified shearing-box simulations mainly serve for calibrating and interpreting
stratified simulation results. Therefore, we do not aim at a thorough parameter study,
but mainly focus on parameter regimes relevant to real PPDs. In this regard, we
consider the following set of parameters:
\begin{itemize}
\item The Hall length $l_H=0.1H$ or $0.3H$.

\item Net vertical field strength, with $\beta_0=10^4$ and $10^5$.

\item Magnetic field polarity, $B_{z0}>0$ or $B_{z0}<0$.

\item The value of $Am=1$, occasionally 10 and 100.
\end{itemize}

Our simulations use fixed box size of $4H\times4H\times2H$ in ($x$, $y$, $z$)
dimensions. Note that our simulation box height is $2H$ rather than $H$ typically used
in unstratified shearing-box simulations, which has the potential to accommodate larger
spatial structures while not being unrealistically tall for real disks. Our unstratified
simulations can be performed with relatively high spatial resolution, $48$ cells per $H$
in the $x-z$ plane (24 in the $y$ dimension). We can not afford the same resolution for
our stratified runs in Sections 3-5, therefore, we also conduct simulations with half the
resolution to justify the use of lower resolution in our stratified simulations.

We have chosen the value of $Am=1$ appropriate for the midplane region of the outer
disk. From Equation (\ref{eq:AmlH}), the Hall length of $l_H\sim0.1$ to $0.3H$ applies to
the range of $R\sim20$ to $50$ AU. Given $\beta_0=10^4$ and $10^5$, the corresponding
value of $\chi_0$ ranges from $0.015$ to $0.14$.

\begin{figure}
    \centering
    \includegraphics[width=90mm]{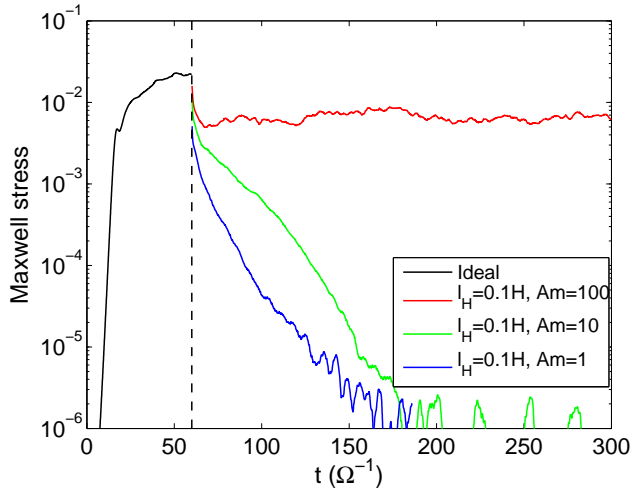}
  \caption{Non-linear sustainability of the MRI turbulence in the case of $B_{z0}<0$.
  The run is initialized with ideal MHD with $\beta_0=10^4$ till $t=60\Omega^{-1}$ before the
  Hall (with $l_H=0.1H$) and AD terms are turned on. Without linearly unstable MRI mode,
  turbulence is sustained for $Am=100$ but decays for $Am=10$ and $1$.}\label{fig:histneg}
\end{figure}

For $B_{z0}<0$, and for this range of $\chi_0$ there is no linearly unstable MRI mode.
However, this does not necessarily relate to the non-linear sustainability, given the
relatively small value of $l_H$. Therefore, in our simulations, we first run the
simulations in the ideal MHD limit to time $t=60\Omega^{-1}$, then turn on non-ideal MHD
terms and evolve further to time $t=300\Omega^{-1}$. In Figure \ref{fig:histneg} we show the
time evolution of two runs in the case of $B_{z0}<0$, with fixed $l_H=0.1$, $\beta_0=10^4$
but different $Am=1$, $10$ and $100$. We see that for $Am=100$, MRI turbulence can be
sustained but at a lower level, while for $Am=10$ and $1$, turbulence is suppressed. We
have tested with other values of $\beta_0$ and $l_H$, and find that as long as $Am=1$, no
sustained MRI turbulence is possible. This implies that under this configuration, the
midplane region of the outer disk is likely the exact analog of the conventional ``dead zone".

\begin{table*}
\caption{List of Unstratified Simulation Runs with $B_{z0}>0$.}\label{tab:unstratruns}
\begin{center}
\begin{tabular}{cccccccccccc}\hline\hline
 Run & Res. & $Am$ & $l_H$ &  $\beta_0$ & $\chi_0$ & $E_k$ & $E_M$ & $\alpha_{\rm Rey}$ & $\alpha_{\rm Max}$ & $\alpha$ & $\alpha_{\rm mag}$ \\\hline
Q3A1B4-R24 & 24 & $1$ & $0.3$ & $10^4$ & $0.047$ & $4.6\times10^{-2}$ & $2.4\times10^{-3}$ & $3.0\times10^{-4}$ & $5.0\times10^{-4}$ & $8.0\times10^{-4}$ & 0.21 \\
Q3A1B4-R48 & 48 & $1$ & $0.3$ & $10^4$ & $0.047$ & $3.1\times10^{-2}$ & $3.6\times10^{-3}$ & $3.8\times10^{-4}$ & $7.8\times10^{-4}$ & $1.2\times10^{-3}$ & 0.22 \\\hline
Q3A1B5-R24 & 24 & $1$ & $0.3$ & $10^5$ & $0.015$ & $1.4\times10^{-2}$ & $3.7\times10^{-3}$ & $4.3\times10^{-4}$ & $8.8\times10^{-4}$ & $1.3\times10^{-3}$ & 0.24 \\
Q3A1B5-R48 & 48 & $1$ & $0.3$ & $10^5$ & $0.015$ & $1.4\times10^{-2}$ & $4.2\times10^{-3}$ & $5.1\times10^{-4}$ & $9.8\times10^{-4}$ & $1.5\times10^{-3}$ & 0.23\\\hline
Q1A1B4-R24 & 24 & $1$ & $0.1$ & $10^4$ & $0.14$ & $1.6\times10^{-2}$ & $2.6\times10^{-3}$ & $5.9\times10^{-4}$ & $6..1\times10^{-4}$ & $1.2\times10^{-3}$ & 0.24 \\
Q1A1B4-R48 & 48 & $1$ & $0.1$ & $10^4$ & $0.14$ & $1.7\times10^{-2}$ & $4.9\times10^{-3}$ & $9.3\times10^{-4}$ & $1.2\times10^{-3}$ & $2.1\times10^{-3}$ & 0.25\\\hline
Q1A1B5-R24 & 24 & $1$ & $0.1$ & $10^5$ & $0.045$ & $1.0\times10^{-2}$ & $1.7\times10^{-3}$ & $3.8\times10^{-4}$ & $2.3\times10^{-4}$ & $6.1\times10^{-4}$ & 0.14 \\
Q1A1B5-R48 & 48 & $1$ & $0.1$ & $10^5$ & $0.045$ & $9.8\times10^{-3}$ & $1.1\times10^{-3}$ & $3.9\times10^{-4}$ & $2.3\times10^{-4}$ & $6.1\times10^{-4}$ & 0.20\\\hline
\end{tabular}
\end{center}
$l_H$ is normalized to $H$, $E_k$ and $E_M$ are normalized to midplane gas pressure $\rho_0c_s^2$.
See Section 2.3 for details.
\end{table*}

For $B_{z0}>0$, the background field configuration is unstable to the MRI. We provide the list
of runs and diagnostic quantities in Table \ref{tab:unstratruns}. The runs are named in
the form of Q$x$A$y$B$z$-R$w$, where $x=10l_H/H$, $y=$Am, $z=\log_{10}(\beta_0)$, and
$w$ is the numerical resolution (24 or 48 per $H$). In all cases, we have fixed the value of
$Am=1$. We find that vigorous turbulence is quickly developed for all runs. Many of these runs
show secular effects in their evolution (to be discussed later), hence we run these simulations
for very long time to $t=1440\Omega^{-1}$ and extract turbulence statistics by performing time
and volume averages
after $t=1120\Omega^{-1}$ (denoted by the over line). Major diagnostic quantities include the
kinetic energy density $E_k=\overline{\rho v^2/2}$, magnetic energy density $E_M=\overline{B^2}/2$,
the Maxwell stress $\alpha^{\rm Max}\equiv-\overline{B_xB_y}$ and Reyholds stress
$\alpha^{\rm Rey}\equiv\overline{\rho v_xv_y}$ (normalization $\rho_0c_s^2$ is omitted since it
equals 1 in code unit). The total Shakura-Sunyaev $\alpha$ is $\alpha^{\rm Max}+\alpha^{\rm Rey}$.
Another useful diagnostic is $\alpha^{\rm mag}\equiv\alpha^{\rm Max}/E_M$ (e.g.
\citealp{Hawley_etal11,Sorathia_etal12}), which is considered as a useful indicator for numerical
convergence.

\begin{figure}
    \centering
    \includegraphics[width=90mm]{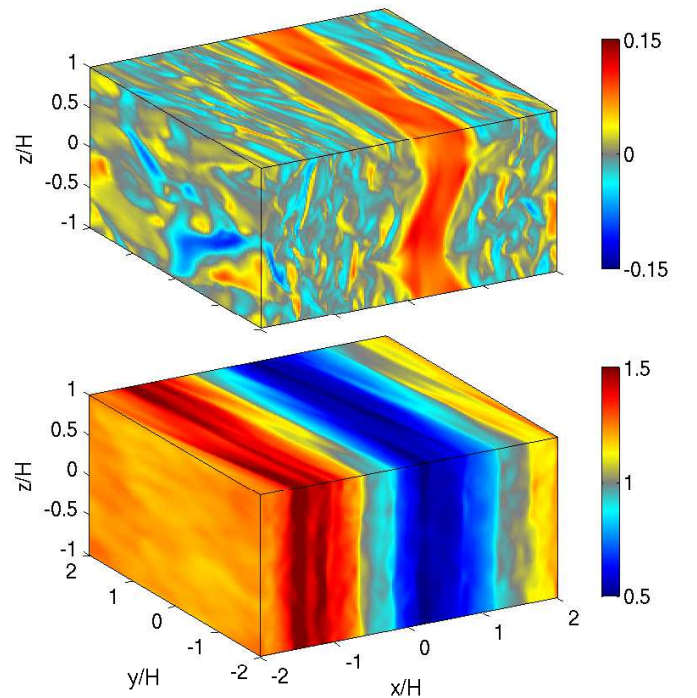}
  \caption{Snapshot from the end of our unstratified run Q3A1b4-R48 with $Am=1$,
  $l_H=0.3H$ and $B_{z0}>0$. The top panel shows the vertical magnetic field
  $B_{z0}$, and the bottom panel shows the gas density $\rho$.
  }\label{fig:zonal}
\end{figure}

First, we find that for relatively large $l_H=0.3H$, and relatively strong field $\beta_0=10^4$,
strong zonal field \citep{KunzLesur13} is gradually built up on relatively long timescales ($\sim100$
orbits), which results from concentration of vertical magnetic flux pertaining to the Hall effect. In
Figure \ref{fig:zonal}, we show the final snapshot of our run Q3A1B4-R48 at time
$t=1440\Omega^{-1}$, which clearly shows the zonal field structure. On the other hand, we find
that the zonal field coexists with vigorous turbulence, and gives an $\alpha$ value of $\sim10^{-3}$.
The presence of vigorous turbulence, rather than remaining in the ``low-transport state", is largely
due to relatively strong magnetic diffusion with $Am=1$, which acts against the buildup of magnetic
flux as discussed in \citet{KunzLesur13}. We do not observe such prominent zonal field structures
in other runs with smaller $l_H$ and weaker magnetic fields.

In the mean time, we find that in essentially all of our unstratified simulations, density variation also
show significant zonal structure, leading to strong zonal flows to balance the pressure gradient
of the zonal density variation \citep{Johansen_etal09}. Such density variation is not captured in
\citet{KunzLesur13} due to their usage of incompressible code. The density variation for our run
Q3A1B4-R48 is shown in the bottom panel of Figure \ref{fig:zonal}, which exhibits excessive density
variation of $\sim50\%$. As a result, the kinetic energy displayed in Table \ref{tab:unstratruns} is
largely dominated by the kinetic energy associated with the zonal flow ($v_y\sim0.2-0.3c_s$). Other
runs develop weaker zonal density variations, and weaker zonal flows as well, which take place over
more than 100 orbital timescale and show secular variations. Full discussion on such zonal flows is
beyond the scope of this paper, but phenomenologically, we observe that stronger zonal flow is
launched for larger $l_H$ and stronger background field from our unstratified simulations.

In all our simulations, sustained MRI turbulence at the level of $\alpha\sim10^{-3}$ is
obtained. Stronger background vertical field leads to stronger turbulence, and larger
$l_H$ also leads to stronger turbulence until the zonal field configuration is developed,
where turbulence level is reduced.
We caution that for the parameters considered here, the most unstable MRI mode is
not well resolved. For best resolved case (run Q3A1B4-R48), we find from Equation
(\ref{eq:lambdam}) that the most unstable wavelength amounts to about $13$ cells.
We do not expect our simulations to show unambiguous convergence on the value
of $\alpha$ (and in fact the value of $\alpha$ is also affected by the development of
the zonal flows, which show long timescale variations). Nevertheless, by looking
at the value of $\alpha_{\rm mag}$, we find that the low and high resolution simulations
give consistent values for all cases except for run Q1A1B5. Moreover, by inspecting
the snapshots in runs with different resolutions, we find their evolutionary behaviors are
qualitatively similar in all cases. This gives us confidence that $24$ cells per $H$
adopted in our stratified runs is sufficient to capture the of essential properties of the
MRI in the Hall-AD regime.

In sum, our unstratified simulations of the MRI in the presence of both the Hall effect
and AD indicate that under conditions appropriate for the outer region of PPDs
($Am\sim1$), MRI can not be self-sustained in the midplane if $B_{z0}<0$, while for
$B_{z0}>0$, the self-sustained turbulence always exists at the level of $\alpha\sim10^{-3}$.
We find zonal fields when the Hall term and background field is relatively strong, and find
zonal flows develop in all cases.

\section[]{Setup of 3D Stratified Simulations}\label{sec:3dsetup}

We perform a series of 3D stratified shearing-box simulations where all non-ideal
MHD effects are included self-consistently. The set up of the simulations follow closely
to those in paper I, with formulation given in his Section 2.1-2.2 and methodology given
in Section 3.1. In brief, we consider a MMSN disk. At a given radial location $R$,
we produce a diffusivity table based on equilibrium chemistry using the chemical reaction
network developed in our earlier works \citep{BaiGoodman09,Bai11a} and the latest
version of the UMIST database \citep{UMIST12}. Dust grains of $0.1\mu$m in size and
abundance of $10^{-4}$ is assumed\footnote{We find that using the complex chemical
reaction network, the resulting ionization fraction in low density and low temperature
regions is, surprisingly, higher than the grain-free case (the same does not hold when
considering the simple network of \citet{OD74}). Since this occurs mainly in the
FUV-dominated surface layer of the outer disk ($\gtrsim30$ AU) where the gas behaves
in the ideal MHD regime, our simulation results are insensitive to this fact. For
consistency we also produce a diffusivity table with grain-free chemistry and choose the
one with higher diffusivity in the final table.}. Standard sources of ionization including
cosmic rays, X-rays and radioactive decay are included. We further include an effective
treatment of the far-UV (FUV) ionization which substantially reduces non-ideal MHD effects
toward disk surface, calibrated with the models of \citet{Walsh_etal10,Walsh_etal12}.
The gas essentially behaves in the ideal MHD regime in the FUV ionization layer.
The diffusivities have the form $\eta_O$, $\eta_H\propto B$ and $\eta_A\propto B^2$,
which is applicable given the small grain abundance.

Unlike in paper I, simulations in this work are full-3D, since we expect the development
of MRI turbulence. All our simulations have vertical domain extending from $z=-6H$ to
$6H$ using a resolution of 24 cells per $H$ in $x$ and $z$, and half the resolution in
$y$. A density floor of $5\times10^{-6}\rho_0$ is applied for all simulations to avoid
numerical difficulties in the strongly magnetized disk surface region (where $\rho_0=1$
is the midplane gas density in code unit). For simulations in Section 4 (at $R=30$ AU),
we use very extended horizontal box size of $6H\times12H$ in ($x, y$) to better
accommodate potentially large-scale structures. Note that for MMSN disk at 30 AU,
the disk aspect ratio $H/R\approx0.078$, hence the radial box size $\sim14$ AU,
which is about the maximum size where shearing-sheet approximation can be
considered as reasonable.  Smaller horizontal domain size of $4H\times8H$ is used
for simulations in Sections 5-6 to reduce computational cost.

All simulations are started with all non-ideal MHD terms turned on, and are initialized
with uniform vertical magnetic field $B_{z0}$ characterized by midplane plasma $\beta_0$,
together with a sinusoidally varying (in $x$) vertical field $B_{z1}$ to avoid strong initial
channel flows \citep{BaiStone13a}. To allow the simulations to saturate quickly, we choose
the amplitude of $B_{z1}$ to be four times $B_{z0}$, and four wavelength of the sinusoidal
variations in $x$:
\begin{equation}
B_z=B_{z0}+4B_{z0}\sin{\bigg(\frac{4\times2\pi x}{L_x}\bigg)}
\end{equation}
Simulations are typically run for about 153 orbits to $t=960\Omega^{-1}$ or about 115
orbits to $t=720\Omega^{-1}$.

We have slightly modified the vertical outflow boundary condition compared with paper I.
Here, the boundary condition assumes hydrostatic equilibrium in $\rho$, outflow in $v_z$,
zero gradient in $B_z$, $v_x$ and $v_y$ (same as paper I), while $B_x$ and $B_y$ are
reduced proportionally as density in the ghost zones (different from paper I, same as in
\citealp{Simon_etal13b}). We do observe that the evolution of mean magnetic fields
somewhat depends on the treatment of the outflow boundary condition, which reflects
the limitation of shearing box when using open boundaries in the presence of disk outflow.
Some of its influences will be discussed in the main text. Nevertheless, the general
properties of the flow do not sensitively depend on the choice of vertical boundary
condition \citep{Fromang_etal13}.

We consider disk radii of $R=5$ AU, $15$ AU, $30$ AU and $60$ AU, where at each
radius we consider $\beta_0=10^4$ and $10^5$, and for both magnetic polarities.
We mainly focus on two disk radii: $R=30$ AU (Section 4), where we further conduct
Hall-free simulations for detailed comparison; and $R=5$ AU (Section 5), where
comparisons with quasi-1D simulations in paper I will be made. All 3D simulations are
listed in Table \ref{tab:stratruns}, and each run is named as R$x$b$y$H$*$, where $x$
represents disk radius in AU, $y=log_{10}\beta_0$, and $*$ can be $0$, `$+$' or `$-$'
for simulations excluding the Hall term ($0$), with the Hall term and $B_{z0}>0$, with the
Hall term and $B_{z0}<0$.

\section[]{Simulation Results: 30 AU}\label{sec:30AU}

We focus on $R=30$ AU in this section. We choose this radius because we find that
at this location, the Hall effect around disk midplane is about equally important as AD.
The disk is likely to develop more stable configurations at smaller disk radii (for
$B_{z0}>0$) as found in paper I, while the Hall effect becomes less prominent toward
larger radii. This location has been explored in \citet{Simon_etal13a,Simon_etal13b},
where only AD was taken into account with fixed profile of $Am=1$ near the midplane.
Our new simulations self-consistently take into account the ionization-recombination
chemistry, together with the inclusion of the Hall effect.

\begin{figure*}
    \centering
    \includegraphics[width=180mm]{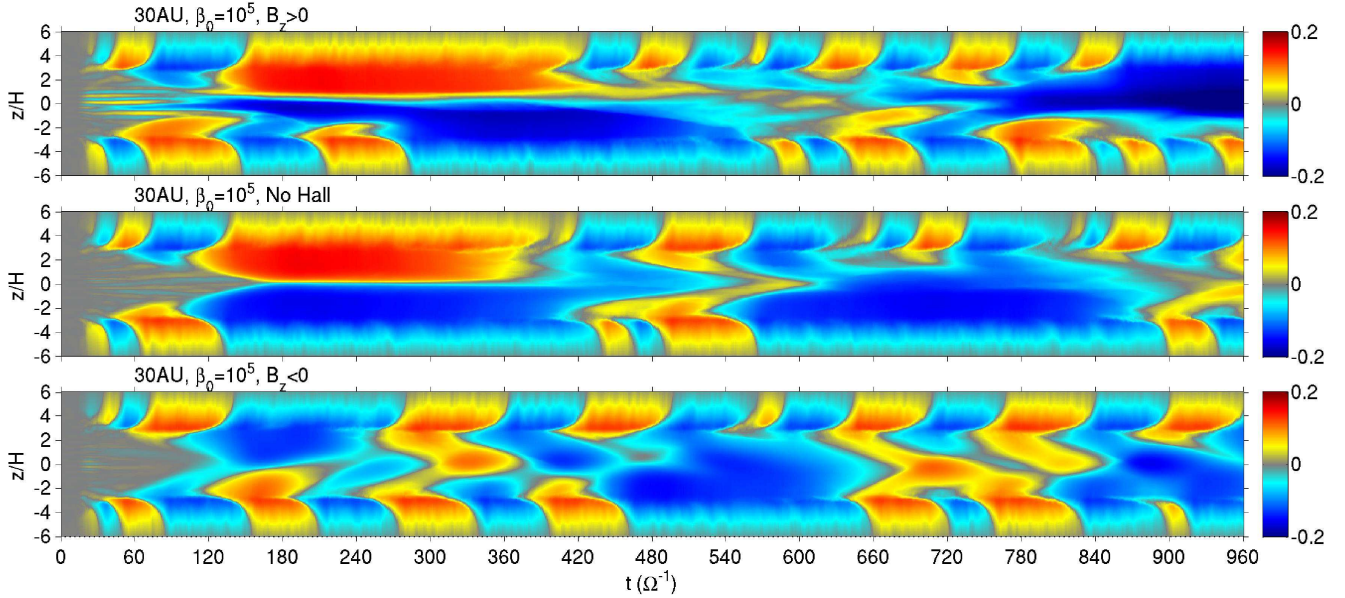}
  \caption{The time evolution for the vertical profile of horizontally averaged $B_y$ in
  our runs at 30 AU with $\beta_0=10^5$. The top, middle and bottom panels
  correspond to runs R30b5H+, R30b5H0 and R30b5H$-$, i.e., Hall turned on with
  $B_{z0}>0$, Hall-free, and Hall turned on with $B_{z0}<0$.
  }\label{fig:30AUb5hist}
\end{figure*}

\begin{figure*}
    \centering
    \includegraphics[width=180mm]{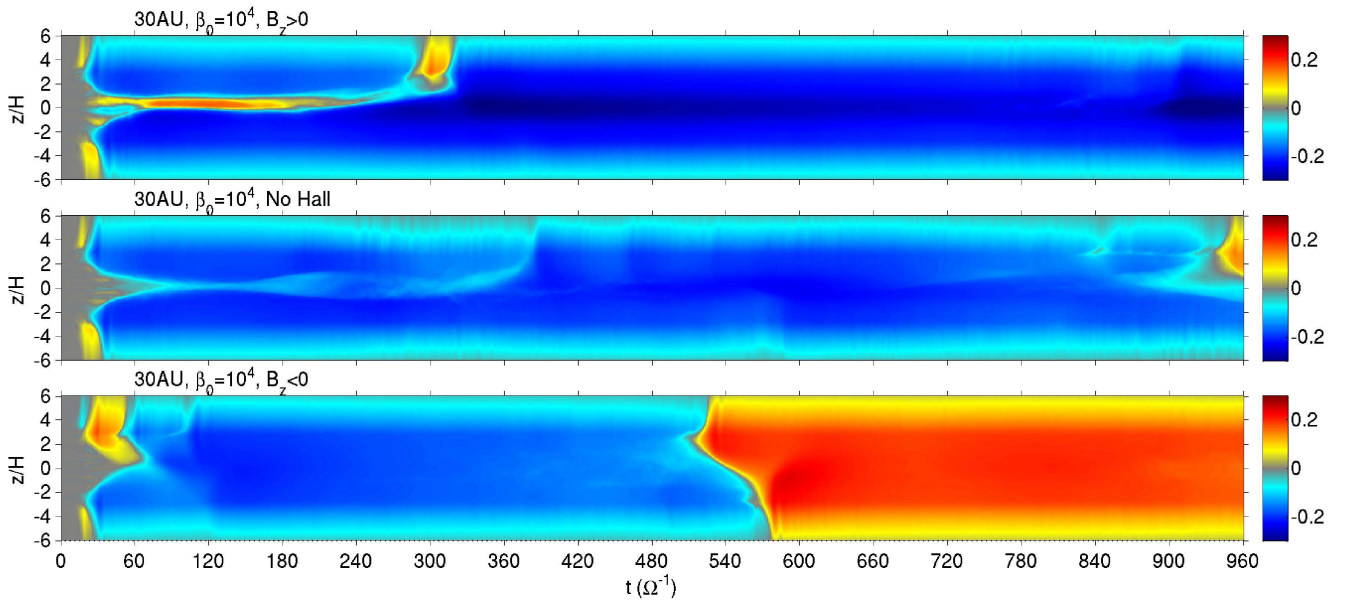}
  \caption{Same as Figure \ref{fig:30AUb5hist}, but for runs at 30 AU with $\beta_0=10^4$.
  }\label{fig:30AUb4hist}
\end{figure*}

We perform a total of 6 simulations with $\beta_0=10^5$ and $10^4$. All these runs
lead to vigorous MRI turbulence in the surface layer, and in the presence of net vertical
magnetic field, they always launch outflows. Different aspects of these simulations are
discussed in the subsections below.

\begin{table*}
\caption{List of Stratified Simulation Runs.}\label{tab:stratruns}
\begin{center}
\begin{tabular}{ccccccccccccc}\hline\hline
 Run & R (AU) & Hall? & $B_{z0}$ &  $\beta_0$ & Box size (H) & $T$ ($\Omega^{-1}$) &
 $\alpha^{\rm Max}$ & $\alpha^{\rm Rey}$ & $\delta v_z$ & $\dot{M}_{\rm out}$ &
 $|T_{z\phi}^{\rm Max}|$ & Section \\\hline\hline
R5b5H+ & 5 & Yes & $+$ & $10^5$ & $4\times8\times12$ & 360 & $6.5\times10^{-3}$ &
                    $1.6\times10^{-5}$ & $1.0\times10^{-2}$ & $3.0\times10^{-4}$ & $3.1\times10^{-4}$ & 5\\\hline
R5b5H-- & 5 & Yes & $-$ & $10^5$ & $4\times8\times12$ & 360 & $4.2\times10^{-4}$ &
                    $3.4\times10^{-5}$ & $5.0\times10^{-3}$ & $1.1\times10^{-4}$ & $1.64\times10^{-4}$ & 5\\\hline
R5b4H-- & 5 & Yes & $-$ & $10^4$ & $4\times8\times12$ & 360 & $1.3\times10^{-3}$ &
                    $4.6\times10^{-4}$ & $1.4\times10^{-3}$ & $3.2\times10^{-4}$ & $6.7\times10^{-4}$ & 5\\\hline\hline

R15b5H+ & 15 & Yes & $+$ & $10^5$ & $4\times8\times12$ & 720 & $2.3\times10^{-3}$ &
                   $1.3\times10^{-4}$ & $5.2\times10^{-3}$ & $2.9\times10^{-4}$ & $2.5\times10^{-4}$ & 6.1\\\hline
R15b5H-- & 15 & Yes & $-$ & $10^5$ & $4\times8\times12$ & 720 & $7.2\times10^{-4}$ &
                   $2.8\times10^{-5}$ & $2.4\times10^{-3}$ & $2.3\times10^{-4}$ & $2.4\times10^{-4}$ & 6.1\\\hline
R15b4H+ & 15 & Yes & $+$ & $10^4$ & $4\times8\times12$ & 720 & $2.3\times10^{-3}$ &
                   $3.1\times10^{-4}$ & $8.2\times10^{-3}$ & $6.1\times10^{-4}$ & $8.8\times10^{-4}$ & 6.1\\\hline
R15b4H-- & 15 & Yes & $-$ & $10^4$ & $4\times8\times12$ & 720 & $3.0\times10^{-3}$ &
                   $1.7\times10^{-4}$ & $8.6\times10^{-3}$ & $8.0\times10^{-4}$ & $1.1\times10^{-3}$ & 6.1\\\hline\hline

R30b5H+ & 30 & Yes & $+$ & $10^5$ & $6\times12\times12$ & 960 & $1.9\times10^{-3}$ &
                    $3.9\times10^{-4}$ & $2.0\times10^{-2}$ & $2.2\times10^{-4}$ & $2.1\times10^{-4}$ & 4\\\hline
R30b5H0 & 30 & No & $+$ & $10^5$ & $6\times12\times12$ & 960 &$1.5\times10^{-3}$ &
                    $2.9\times10^{-4}$ & $1.5\times10^{-2}$ & $2.3\times10^{-4}$ & $2.2\times10^{-4}$ & 4\\\hline
R30b5H-- & 30 & Yes & $-$ & $10^5$ & $6\times12\times12$ & 960 & $1.4\times10^{-3}$ &
                    $2.2\times10^{-4}$ & $1.3\times10^{-2}$ & $2.3\times10^{-4}$ & $2.2\times10^{-4}$ & 4\\\hline\hline

R30b4H+ & 30 & Yes & $+$ & $10^4$ & $6\times12\times12$ & 960 & $6.1\times10^{-3}$ &
                    $4.4\times10^{-4}$ & $2.0\times10^{-2}$ & $1.5\times10^{-3}$ & $1.7\times10^{-3}$ & 4\\\hline
R30b4H0 & 30 & No & $+$ & $10^4$ & $6\times12\times12$ & 960 & $4.8\times10^{-3}$ &
                    $5.4\times10^{-4}$ & $2.4\times10^{-2}$ & $1.1\times10^{-3}$ & $1.4\times10^{-3}$ & 4\\\hline
R30b4H-- & 30 & Yes & $-$ & $10^4$ & $6\times12\times12$ & 960 & $5.0\times10^{-3}$ &
                    $6.5\times10^{-4}$ & $2.4\times10^{-2}$ & $1.2\times10^{-3}$ & $1.4\times10^{-3}$ & 4\\\hline\hline

R60b5H+ & 60 & Yes & $+$ & $10^5$ & $4\times8\times12$ & 720 & $2.9\times10^{-3}$ &
                   $5.7\times10^{-4}$ & $2.5\times10^{-2}$ & $2.4\times10^{-4}$ & $2.2\times10^{-4}$ & 6.2\\\hline
R60b5H-- & 60 & Yes & $-$ & $10^5$ & $4\times8\times12$ & 720 & $2.6\times10^{-3}$ &
                   $5.0\times10^{-4}$ & $2.1\times10^{-2}$ & $2.4\times10^{-4}$ & $2.1\times10^{-4}$ & 6.2\\\hline
R60b4H+ & 60 & Yes & $+$ & $10^4$ & $4\times8\times12$ & 720 & $9.3\times10^{-3}$ &
                   $4.4\times10^{-4}$ & $8.9\times10^{-3}$ & $2.0\times10^{-3}$ & $1.9\times10^{-3}$ & 6.2\\\hline
R60b4H-- & 60 & Yes & $-$ & $10^4$ & $4\times8\times12$ & 720 & $7.3\times10^{-3}$ &
                   $4.3\times10^{-4}$ & $1.1\times10^{-2}$ & $1.8\times10^{-3}$ & $2.0\times10^{-3}$ & 6.2\\\hline\hline
\hline
\end{tabular}
\end{center}
Note: $\alpha^{\rm Max}$ and $\alpha^{\rm Rey}$ are computed within $z=\pm4.5H$,
$T_{z\phi}^{\rm Max}$ is evaluated at $z=\pm4.5H$,
and $\delta v_z$ is the turbulent vertical velocity within $z=\pm2H$.
See Section 4 for details.
\end{table*}

\subsection[]{Evolution of Large-scale Toroidal Field}\label{ssec:evolve}

Global evolution of the system is largely controlled by magnetic fields, hence
we first discuss the overall evolution of large-scale toroidal field from our simulations
as a standard diagnostic. Starting from runs with $\beta_0=10^5$: R30b5H+, R30b5H0
and R30b5H$-$, we show in Figure \ref{fig:30AUb5hist} the time evolution of horizontally
averaged $B_y$ for all three runs. Since the initial conditions for these simulations are
identical (except for negative sign of $B_{z0}$ for run R30b5H$-$), these runs initially
proceed in a similar way. The Hall and AD terms become progressively more important
as (midplane) magnetic fields become stronger and the three runs then evolve differently.
All three cases show prominent level of dynamo activities emanating from the surface
layer, where the sign of mean $B_y$ alternates over time. The alternation behavior is quite
irregular, and to some extent similar to ideal MHD simulations with modestly strong vertical
magnetic flux ($\beta_0\gtrsim10^3$, \citealp{BaiStone13a}), which contrasts with the
conventional MRI dynamo (zero net vertical magnetic field in ideal MHD) with very periodic
cycles of about $10$ orbits (e.g., \citealp{Davis_etal10,Shi_etal10}).

We next discuss simulations with $\beta_0=10^4$, with three runs R30b4H+,
R30b4H0 and R30b4H$-$. Similar to the weaker field case, all three runs develop
vigorous turbulence mainly in the surface layer due to FUV ionization (see next subsection).
The time evolution of horizontally averaged $B_y$ for the three runs is shown in
Figure \ref{fig:30AUb4hist}. We see that the MRI dynamo is suppressed in all cases and
the mean toroidal field is predominantly one sign. This is generally a consequence of
stronger net vertical field, which is an analog of the ideal MHD case \citep{BaiStone13a}.
While the system is turbulent, toroidal field is always the dominant field component, and
when the dynamo is suppressed, this field component is dominated by the mean field.
Therefore, the space-time plot of mean $B_y$ largely characterizes the evolution of the
system. However, by viewing individual simulation snapshots, localized patches
possessing opposite sign of toroidal field do exist in runs R30b4H0 and R30b4H$-$. In the
latter case, the region with opposite $B_y$ gradually grows and eventually leads to the
reversal of mean toroidal field in the disk (bottom panel of the Figure). We have continued
this run further and found that the mean $B_y$ will reverse again after another $\sim50$
orbits, and this cycle is likely to continue. Similarly, positive $B_y$ region started to
dominate the upper half of the disk near the end of our run R30b4H0.

The secular evolution of the mean field discussed above exists in all our simulations to a
certain extent, which is partly related to the limitations of the shearing-box approach:
due to the imposed net vertical field which presumably connects to infinity, the mean field
in the disk should be in causal contact with the field beyond, but the causal connection is
truncated with prescribed outflow boundary condition.
Since most activities in the disks are magnetically-driven, the secular evolution of the
mean fields also makes the level of turbulence in the disks time variable. For example, in
run R30b4H$-$, the midplane region exhibits stronger turbulent activities around time
$t=480-600\Omega^{-1}$ with turbulent velocity about a factor of 3 higher than some other
periods. Therefore, the readers should bear in mind about the potential uncertainties due
to such variabilities. To obtain the vertical profiles of various diagnostic quantities in the
next subsection, we
will perform time average for around $75-100$ orbits, expecting relatively long-term averages
to provide reasonably realistic mean values.

\subsection[]{Stress Profiles and Level of Turbulence}

\begin{figure*}
    \centering
    \includegraphics[width=150mm]{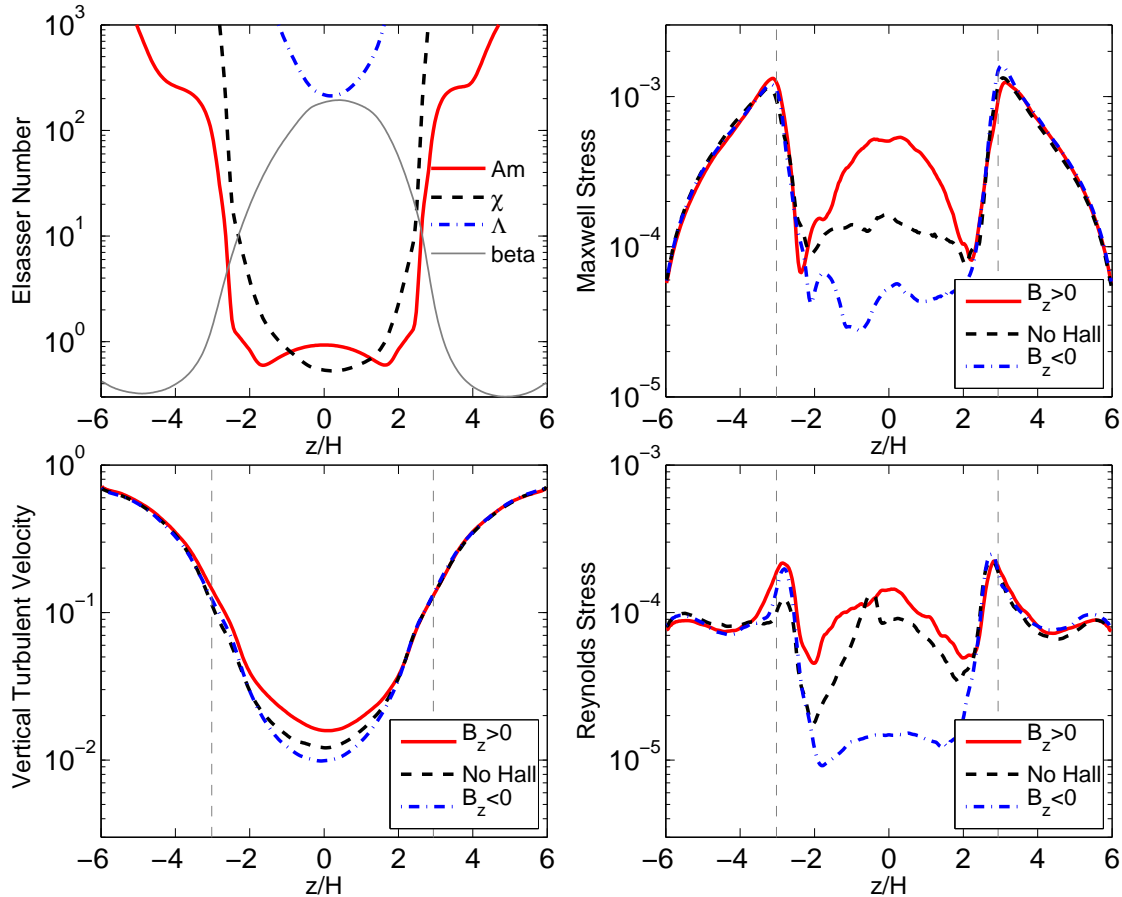}
  \caption{The vertical profiles of various horizontally averaged diagnostic quantities
  from our runs at 30 AU with $\beta_0=10^5$. Top left: the Ohmic ($\Lambda$), Hall
  ($\chi$) and ambipolar ($Am$) Elsasser numbers in blue dash-dotted, black dashed
  and red solid lines, together with plasma $\beta$ in thin gray line. The profile is
  extracted from the Hall-free run R30b5H0 (almost identical to the other two runs).
  Bottom left: vertical turbulent velocity. The rest three panels show various profiles
  fro all three runs of R30b5H+ (red solid), R30b5H0 (black dashed) and R30b5H$-$
  (blue dash-dotted). Top right: Maxwell stress $-B_xB_y$. Bottom right: Reynolds
  stress $\rho v_xv_y$. The gray vertical dashed lines mark the location where
  $Am=100$ in run R30b5H0.}\label{fig:30AUb5prof}
\end{figure*}

\begin{figure*}
    \centering
    \includegraphics[width=180mm]{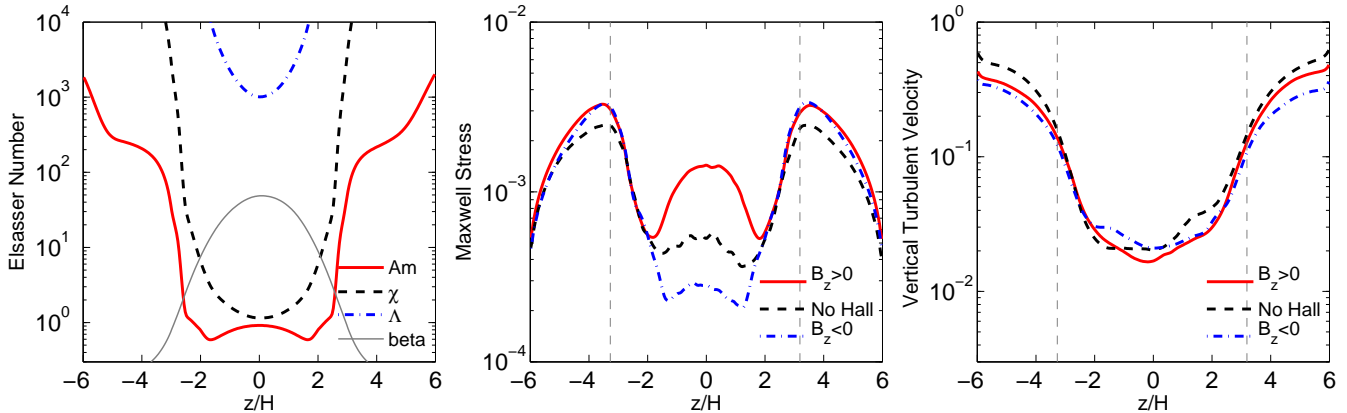}
  \caption{Same as the Figure \ref{fig:30AUb5prof} without the bottom right panel,
  but for runs at 30 AU with $\beta_0=10^4$. The vertical dashed line labels the
  location where $Am=100$ in run R30b4H0.}\label{fig:30AUb4prof}
\end{figure*}

Based on the time evolution of the mean field, we extract useful diagnostic
quantities and average them in time from $t=480\Omega^{-1}$ onward
for simulations with $\beta_0=10^5$, and from $t=360\Omega^{-1}$ onward
for simulations with $\beta_0=10^4$. In Figures \ref{fig:30AUb5prof} and
\ref{fig:30AUb4prof}, we show the time-averaged vertical profiles of various
diagnostic quantities from these simulations.

The relative importance of various non-ideal MHD effects can be best viewed
from the top left panel of Figure \ref{fig:30AUb5prof} and the left panel of
Figure \ref{fig:30AUb4prof}, which show the profiles of the Elsasser numbers
(based on the Hall-free run in each case, but the runs with Hall term generally
give almost the same profiles).
Clearly, Ohmic resistivity is completely negligible with $\Lambda\gg100$ at all
heights. With $\beta_0=10^5$, both the Hall effect and AD are important within
$z\sim\pm2-2.5H$ with $\chi$ and $Am$ being around 1, and the range of
influence by AD extends higher from the midplane than the Hall effect. The Hall
effect is less important relative to AD with stronger net flux $\beta_0=10^4$
because the resulting total field is stronger.
Beyond $z\sim2.5H$, the FUV ionization catches up and all non-ideal MHD
effects are greatly reduced. Beyond $z=\pm3H$, the gas essentially behaves
in the ideal MHD regime with $Am>100$.

Vigorous MRI turbulence takes place beyond about $z\sim\pm2.5H$ thanks
to FUV ionization. As a result, the profile of the Maxwell stress
$T_{R\phi}^{\rm Max}=-B_xB_y$ peaks at around $z=\pm3H$, as shown in the
top right panel of Figure \ref{fig:30AUb5prof} and middle panel of Figure
\ref{fig:30AUb4prof}. Beyond $z\sim\pm3H$, the Maxwell stress drops because
disk density drops and it enters the magnetically dominated corona (plasma
$\beta<1$). All three runs at a given $\beta_0$ show very similar properties in
this region, since the gas behaves in the ideal MHD regime. Runs with
$\beta_0=10^4$ have systematically higher Maxwell stress than the corresponding
$\beta_0=10^5$ runs by a factor of 3-4 as a result of stronger background field.

The midplane region is where three simulations at fixed $\beta_0$ are expected
differ due to the Hall effect. The most prominent difference lies in the Maxwell stress.
The runs with $B_{z0}>0$ give the highest stress that peaks at the midplane. This
is related to the Hall-shear instability \citep{Kunz08}, which operates only when
$B_{z0}>0$, and is responsible for generating stronger horizontal magnetic fields
hence Maxwell stress in the inner disk (\citealp{Lesur_etal14}, paper I).
Here, the effect is much less prominent than in the inner disk studied in paper I
and \citet{Lesur_etal14} since the Hall effect is only modestly significant
($\chi\sim1$). The runs with $B_{z0}<0$ give the lowest midplane Maxwell
stress, while the Maxwell stress from R30b5H0 (without the Hall term) lies in
between. This is again consistent with the expectation from paper I that
horizontal magnetic field tends to be reduced for negative $B_{z0}$.

As discussed in Section 2, for $B_{z0}>0$, the midplane region is unstable to
the MRI, and the level of the MRI turbulence is expected to be stronger than
the Hall-free case. For $B_{z0}<0$, self-sustained MRI turbulence is not
expected due to the Hall effect. 
To characterize the level of turbulence, we consider the vertical component
of the rms velocity, which are shown in the bottom left panel of Figure
\ref{fig:30AUb5prof} and the right panel of Figure \ref{fig:30AUb4prof} for the
two sets of runs. They are computed based on the turbulent kinetic energy at
each height. In the same way, we define $\delta v_z$ to be the rms vertical
velocity fluctuation within $z=\pm2H$ for all our runs, and have included it in
Table \ref{tab:stratruns}.

We see that the turbulent rms vertical velocity reaches $\sim0.3-0.8c_s$ at
disk surface ($z\sim\pm4H$) for all these runs, while is reduced by more than
one order of magnitude to $\sim0.01-0.03c_s$ around disk midplane. For
$\beta_0=10^5$, the run with $B_{z0}>0$ gives higher midplane turbulent
velocity while the run with $B_{z0}<0$ gives the lowest, and the Hall-free run
lies in between, which is consistent with our expectation. Nevertheless, the
difference is within a factor of $2$, hence the role of the Hall effect in the midplane
turbulent activities is only modest. While we caution that the level of turbulence
in the $B_{z0}>0$ case may be underestimated due to the lack of numerical
resolution, the overall scenario is similar to the Hall-free case, and consistent
with earlier stratified AD simulations of \citep{Simon_etal13b}, where the
midplane region was termed as ``ambipolar-damping" zone (the region MRI
active but with low turbulence level due to AD). In the case of $B_{z0}<0$
where MRI can not be self-sustained at disk midplane, the midplane turbulent
motion is most likely induced by the strong MRI turbulence in the disk surface
layer, which is a direct analog of the {\it conventional} ``Ohmic dead zone" the
inner disk (e.g., \citealp{FlemingStone03,OishiMacLow09}).

For $\beta_0=10^4$, we find that the level of midplane turbulence in all three
runs are very similar (modulo some secular variations not reflected in the
time-averaged plots), despite the marked difference in Maxwell stress. We
have checked that for $B_{z0}>0$, the midplane Maxwell stress is dominated
by contributions from large-scale field ($-\overline{B_x}\overline{B_y}$), while
for $B_{z0}<0$, the midplane Maxwell stress is almost entirely due to turbulent
field. Turbulent contributions of the midplane Maxwell stress from the two runs
R30b4H+ and R30b4H$-$ are in fact similar. We have also checked that for
$\beta_0=10^5$, midplane Maxwell stress is always dominated by turbulent
stress. The low level of turbulence in run R30b4H+ may be considered as a
consequence of the strong mean toroidal field ($\overline{B_y}$), which
dominates the magnetic field strength and tends to suppress turbulent motions
(but see also Section 4.4).

Overall, based on the six simulations with different strengths and polarities of
the net vertical field, it is clear that the Maxwell stress profile (hence radial
transport of angular momentum) is layered. Moreover, it appears that
$\delta v_z\approx0.01-0.02c_s$ is a good proxy for the level of turbulence in
the midplane region of the outer disks, with much stronger turbulence in the
FUV ionization layer at disk surface.

\subsection[]{Angular Momentum Transport and Disk Outflow}

Outflow is always launched in shearing-box simulations in the presence of
net vertical magnetic flux (e.g., \citealp{SuzukiInutsuka09}). While this outflow
may serve as a wind launching mechanism, the kinematics of the outflow is not
well characterized in shearing-box simulations because the rate of the mass
outflow does not converge with simulation box height \citep{Fromang_etal13}
and there are also symmetry issues \citep{BaiStone13a}. Therefore, we do
not aim at fully characterizing the outflow properties, but simply provide some
basic diagnostics for reference. We calculate the rate of mass outflow leaving
the simulation box $\dot{M}_{\rm out}$. It is computed by time averaging the
sum of vertical mass flux at the two vertical boundaries. 
We also calculate the $z\phi$ component of the Maxwell stress tensor
$T_{z\phi}^{\rm Max}=-B_zB_\phi$, which determines the rate of wind-driven
angular momentum transport (if the outflow is eventually incoporated into a
global magnetocentrifugal wind). In the laminar case, $T_{z\phi}$ can be
conveniently evaluated at the base of the wind where the toroidal velocity
transitions from sub-Keplerian to super-Keplerian
\citep{BaiStone13b,Bai13}. Since most of our simulations runs are
highly turbulent at the disk surface, there are ambiguities in defining the base
of the wind (and whether the outflow can become a global wind at all,
\citealp{BaiStone13a}), we simply provide a reference value of time-averaged
$|T_{z\phi}^{\rm Max}|$ evaluated at $z=\pm4.5H$ in Table \ref{tab:stratruns}.

\begin{figure*}
    \centering
    \includegraphics[width=180mm]{zonal_30AUb4.eps}
    \includegraphics[width=180mm]{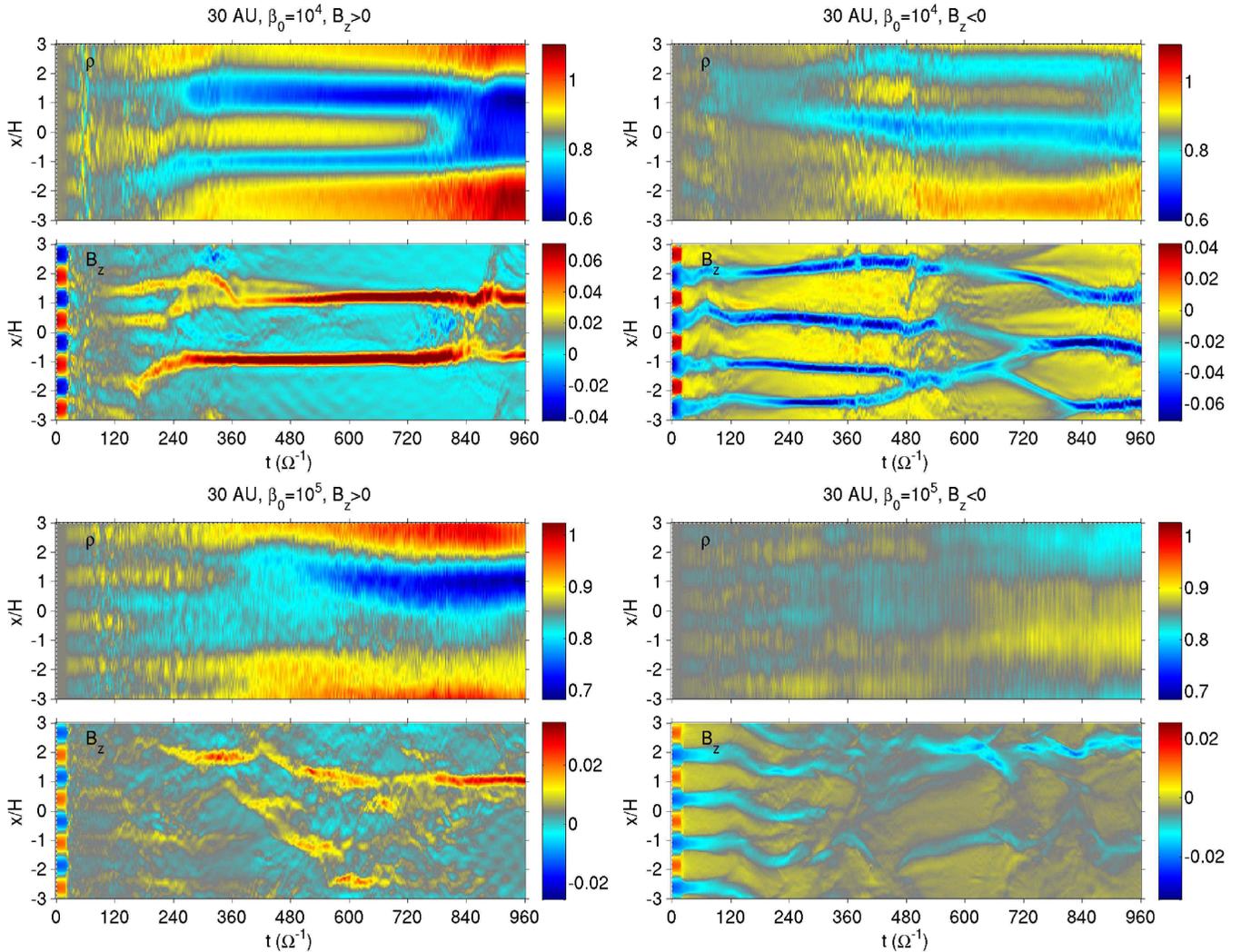}
  \caption{Time evolution of the radial profiles of mean gas density $\rho$ (upper
  panels) and mean vertical magnetic field $B_z$ (lower panels) averaged over the
  $y-z$ plane within $z=\pm2H$ in our runs R30b4H+ (upper left), R30b4H$-$
  (upper right), R30b5H+ (lower left) and R30b5H$-$ (lower right). The color
  scales are centered in their mean values (in code units).}\label{fig:zonal30AU}
\end{figure*}

The value of Shakura-Sunyaev $\alpha$ for stratified disk can be written as
\begin{equation}
\alpha=\frac{\int T_{R\phi}dz}{c_s^2\int\rho dz}\ ,
\end{equation}
where $T_{R\phi}$ has contributions from both the Maxwell stress $(-B_xB_y)$
and Reynolds stress ($\rho v_xv_y$), leading to $\alpha^{\rm Max}$ and
$\alpha^{\rm Rey}$ in Table \ref{tab:stratruns}. From the lower right panel of
Figure \ref{fig:30AUb5hist}, we see that the vertical profile of the Reynolds
stress is generally a factor of several smaller than the Maxwell stress. Due to
uncertainties in characterizing the outflow from shearing-box simulations,
we truncate the vertical integral at $z=\pm4.5H$.
For the six runs, the values of $\alpha$ are found to be around
$1.5-2\times10^{-3}$ for $\beta_0=10^5$ and $5-6\times10^{-3}$ for
$\beta_0=10^4$.

In steady state, the total accretion rate driven from radial transport of angular
momentum (given by $\alpha$) and the putative wind-driven accretion
(given by $T_{z\phi}$) can be approximately written as (e.g., \citealp{Bai13})
\begin{equation}
\begin{split}\label{eq:accret}
\dot{M}&\approx\frac{2\pi}{\Omega}\alpha c_s^2\Sigma+\frac{8\pi}{\Omega}R|T_{z\phi}|\ ,\\
\dot{M}_{-8}&\approx0.82\bigg(\frac{\alpha}{10^{-3}}\bigg)R_{\rm AU}^{-1/2}
+4.1\bigg(\frac{|T_{z\phi}|}{10^{-4}\rho c_s^2}\bigg)R_{\rm AU}^{-3/4}\ ,
\end{split}
\end{equation}
where $R_{\rm AU}$ is the radius measure in AU, and we have assumed
MMSN disk model in the second equation, with $\dot{M}_{-8}$ being accretion
rate measured in $10^{-8}M_{\bigodot}$ yr$^{-1}$.

Using the values from Table \ref{tab:stratruns} with $R=30$ AU, we find that
based on radial transport alone, the resulting accretion rate is about
$0.24-0.33\times10^{-8}M_{\bigodot}$ yr$^{-1}$ for the three runs with
$\beta_0=10^5$ studied here, which is somewhat smaller than desired. If there
were contributions from disk wind, the estimated wind-driven accretion rate is
about $0.7\times10^{-8}M_{\bigodot}$ yr$^{-1}$. The sum of the two contributions
just matches the desired rate of $10^{-8}M_{\bigodot}$ yr$^{-1}$. For
$\beta_0=10^4$, accretion rate resulting from radial angular momentum transport
gives $\sim0.72-0.91\times10^{-8}M_{\bigodot}$ yr$^{-1}$, with potential contribution
from the wind to give $\sim5\times10^{-8}M_{\bigodot}$ yr$^{-1}$.

\subsection[]{Zonal Field and Zonal Flow}

For our 30 AU simulations, we find using Equation (\ref{eq:convert}) and from the
Elsasser number plots in Figures \ref{fig:30AUb5prof} and \ref{fig:30AUb4prof}  that
$l_H\approx0.2H$ around disk midplane, which is about the threshold value to trigger
the zonal field configuration in the unstratified case as discussed in \citet{KunzLesur13}.
In Section 2 we showed in Figure \ref{fig:zonal} that strong zonal field and zonal flow is
observed in unstratified simulations when $\beta_0=10^4$ and $B_{z0}>0$. To check
whether our stratified simulations reveal similar behaviors, we show in Figure
\ref{fig:zonal30AU} the time evolution of mean gas density $\rho$ and $B_z$ for runs
30AUb4H$\pm$ and 30AUb5H$\pm$, averaged in the $y$ and $z$ dimensions, within
the disk region $-2H\leq z\leq2H$.

We find that strikingly, for all runs, vertical magnetic flux is concentrated into thin
(azimuthally extended) shells, while in regions outside these shells, the net vertical
flux is close to zero. In the mean time, there are very prominent radial density
variations characteristic of strong zonal flow. There are clearly secular evolution of
the vertical magnetic flux distribution and zonal flows, which is also related to the secular
behaviors discussed in Section \ref{ssec:evolve}. At first glance, these features appear to
be consistent with those shown in Figure \ref{fig:zonal} from our unstratified simulations.
However, there are distinct differences. In particular, both $B_{z0}>0$ and $B_{z0}<0$
cases show such zonal fields, while from unstratified simulations zonal field is expected
only from the $B_{z0}>0$ case. Also, the width of the zonal field is very small ($<0.5H$),
while from unstratified simulations the width is generally wider than $H$.

In fact, we find that concentration of magnetic flux appears to be a generic behavior
in shearing-box simulations with net vertical magnetic flux. Not only in simulations
with the Hall effect, but our Hall-free simulations at 30 AU, together with many simulations
at other disk radii, all show this behavior to some level. We also find that the concentration
is less prominent when the net vertical field is weaker, as one compares the top and bottom
panels in Figure \ref{fig:zonal30AU}. Accompanied with magnetic flux concentration is the
strong zonal flow, which density variation across the domain up to $\sim30\%$.
Enhanced zonal flow in the presence of net vertical magnetic flux was reported
in \citet{SimonArmitage14} based on stratified shearing-box simulations in the AD
dominated outer disk. Such zonal flows also exist in our earlier simulations including
both Ohmic resistivity and AD further closer in (at 10-20 AU, \citealp{Bai13}),  and we
have verified that in general, there is only one single ``wavelength" of the density/pressure
variation across the radial domain, regardless of the radial domain size (Bai, 2013,
unpublished). From Figure \ref{fig:zonal30AU}, we see that the location where magnetic
flux concentrates significantly correlates with the density minimum. While less evident in
run R30b4H$-$ (the trend weakens in the $B_{z0}<0$ case due to the Hall effect), in
general, the enhanced zonal flow is directly associated with the magnetic flux concentration.

In sum, the zonal field and zonal flow observed in our stratified simulations are not
due to the Hall effect as reported in unstratified simulations, but are correlated
phenomenon generically present in shearing-box simulations with net vertical
magnetic flux. While the saturation of the zonal flow is artificially affected by the
shearing-box since its radial scale is set by the simulation box size, its association
with magnetic flux concentration may make it very likely a physical phenomenon. Our
local simulations here serve as a first study of the PPD gas dynamics including all
non-ideal MHD effects, and it remains to understand their underlying physics and verify
their existence in global simulations.

\section[]{Simulations at 5 AU}

Our second focused location is at relatively small radius of $R=5$ AU, which
compliments our studies in paper I\footnote{To better compare with the results in paper
I, we runs the simulations at 5 AU with the same vertical outflow boundary condition as
paper I instead of the modified version in the rest of the simulations.}. Using quasi-1D
simulations, we have found in paper I that for $B_{z0}>0$, the inner disk launches a
laminar magnetocentrifugal wind which very efficiently drives disk accretion. In
constructing the wind solutions, we enforced reflection symmetry about disk midplane
so that the wind solution has the desired symmetry properties to match to a physical
magnetocentrifugal wind (i.e., horizontal component of the magnetic field must flip
across the disk). It remains to demonstrate that this wind configuration is stable
in 3D without enforcing the symmetry. Another important result from paper I is that for
$B_{z0}<0$, we did not find any stable wind configuration for typically expected level
of vertically magnetic field strength at this location ($\beta_0=10^{5-6}$) since MRI sets
in in a very narrow range of disk height. It remains to demonstrate how the disk behaves
under this situation.

\begin{figure*}
    \centering
    \includegraphics[width=180mm]{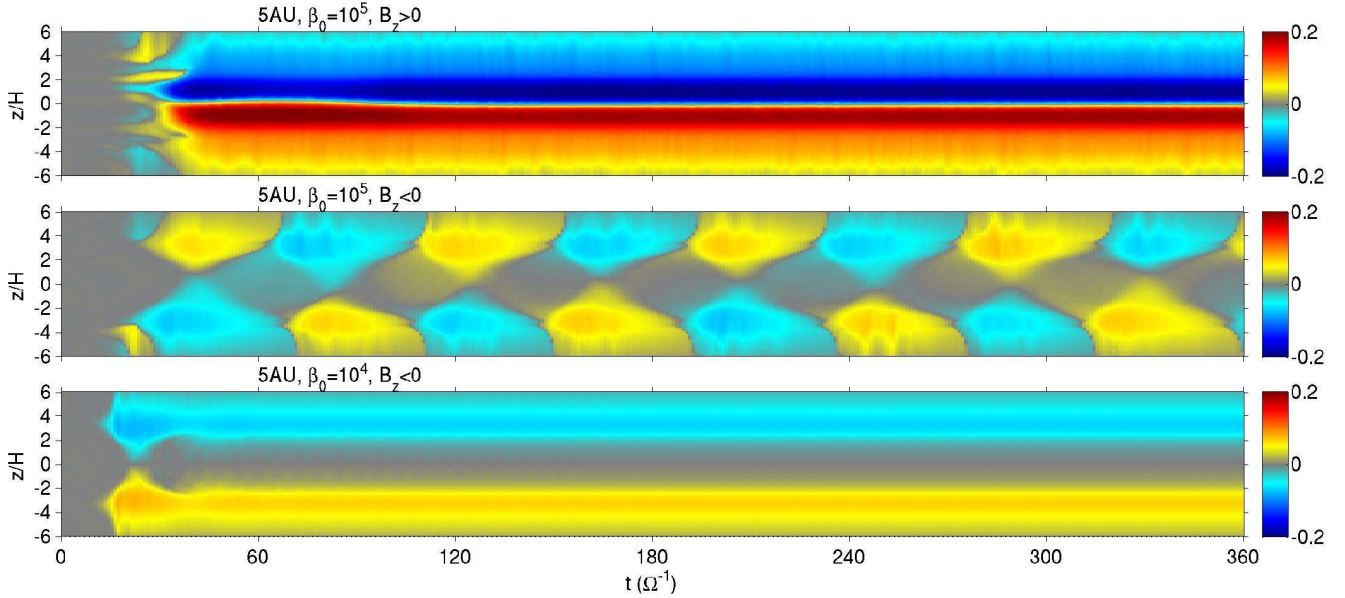}
  \caption{The time evolution for the vertical profile of horizontally averaged $B_y$ in
  our runs at 5 AU. The top, middle and bottom panels correspond to runs R5b5H+,
  R5b5H$-$ and R5b4H$-$.}\label{fig:5AUhist}
\end{figure*}

\begin{figure}
    \centering
    \includegraphics[width=80mm]{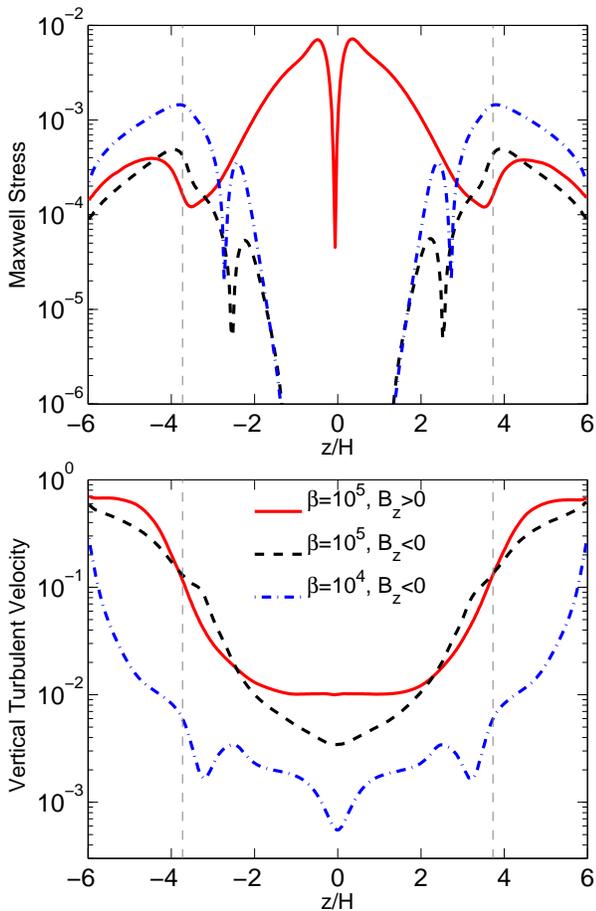}
  \caption{Vertical profiles of Maxwell stress (top) and vertical turbulent velocity
  (bottom) for all three runs at 5 AU, as marked in the legend. The vertical dashed
  line labels the location where $Am=100$ in run R5b5H$-$.
  }\label{fig:5AUprof}
\end{figure}

We have performed three runs. For $B_{z0}>0$ we consider $\beta_0=10^5$ (run
R5b5H+), while for $B_{z0}<0$ we consider $\beta_0=10^5$ and $10^4$ (runs
R5b5H$-$ and R5b4H$-$). From paper I, we expect largely laminar configurations
to be developed for runs R5b5H+ and R5b4H$-$, launching magnetocentrifugal wind;
while the MRI should set in for run R5b5H$-$. In Figure \ref{fig:5AUhist}, we
again show the time evolution of the horizontally averaged $B_y$ in the three runs.
Given the highly regular patterns seen in this Figure, it suffices to run these
simulations just to $t=360\Omega^{-1}$ and perform time average from
$t=180\Omega^{-1}$ onward.

\subsection[]{Simulation with $B_{z0}>0$}

For run R5b5H+, we see from the top panel of Figure \ref{fig:5AUhist} that the system is
able to achieve a largely laminar state as desired. More interestingly, the toroidal field
changes sign almost exactly at the disk midplane, automatically maintaining the reflection
symmetry (more specifically, even-$z$ symmetry, see Figure 9 of \citealp{BaiStone13b}).
Achieving this field geometry is essential for physically launching a magnetocentrifugal
wind, and supports the procedure adopted in paper I where the reflection symmetry across
midplane was enforced. Checking the time-averaged vertical profiles of various quantities,
we find that the result is almost identical with Figure 9 of paper I (with slight difference
since our box extends to $z=6H$ rather than $8H$). For this solution, the horizontal
magnetic field near the midplane is strongly amplified by the Hall shear instability, and the
flip of this horizontal field creates strong current density at the midplane. This contrasts with
the study by \citet{Bai13}, where without the Hall term, the strong current layer was found
to be located offset from the midplane at $z_{SC}\approx1.3H$ in this particular case (see
his Table 2 for run S-R5-b5). It appears that with the inclusion of the Hall term, horizontal
magnetic field tends to flip right across the midplane, rather than from upper layers.

In Figure \ref{fig:5AUprof} we further show the vertical profiles of time-averaged
Maxwell stress and vertical turbulent velocities. For our run R5b5H+, we see that the
Maxwell stress profiles peaks close to disk midplane at rather high level close to
$10^{-2}\rho_0c_s^2$. The dip at midplane is due to the flip of horizontal field, all in
agreement with the results in paper I. For the profile on turbulent velocity, however, we
find that appreciable level of turbulence is present in this run. The turbulent velocity
is again on the order of $0.01c_s$ around the midplane, and increases toward surface
layer at a level very similar to that in the outer disk studied in the previous section. Since
we expect the system to be stable to the MRI, the turbulence mainly originates from
elsewhere: at the midplane, we find that the strong current layer tends to exhibit small
amplitude corrugation from time to time resembling the tearing modes in reconnection
current sheet. Such corrugating motion is likely the source of most random velocities
which propagates toward disk surface layers and becomes amplified due to rapid
density drop.

In sum, for $B_{z0}>0$, our 3D simulation with full box well reproduces the quasi-1D
simulations with enforced reflection symmetry in paper I, and we expect accretion is
mainly driven by magnetocentrifugal wind, together with significant contribution from
radial transport of angular momentum via the large-scale Maxwell stress/magnetic
braking (see Table 2 of paper I). The wind-driven accretion flow mostly proceeds in
the strong current layer where toroidal magnetic field flips \citep{BaiStone13b}, and
here it takes place exactly at disk midplane. Our 3D simulation further reveals the
presence of turbulence, which largely originates from the midplane region where
relatively strong large-scale horizontal magnetic fields flip. The level of turbulence is
similar to that in the outer disk. We also comment that since the system is stable to
the MRI, magnetic flux concentration into thin shells is not observed in this simulation.

\subsection[]{Simulations with $B_{z0}<0$}

For run R5b5H$-$, the system is expected to be unstable to the MRI in a narrow range
of disk height at about $|z|\sim2-3H$. This can roughly be identified from the left panel
of Figure 9 in paper I, where the Hall Elsasser number passes 1 at around $z=2.5H$
and plasma $\beta$ is still not too small (based on the Hall-free run in dashed lines).
Detailed explanation on the onset of the instability is given in Section 5.2 of paper I,
but in brief, it is related to the fact that for $B_{z0}<0$, the Hall term makes the most
unstable MRI wavelength shifts to shorter wavelength when Elsasser number $\chi_0$
is of order unity, allowing the unstable modes to fit into the disk. 
Using full 3D simulations, we see from the middle panel of Figure \ref{fig:5AUhist} that
the large-scale toroidal magnetic field flips
in highly periodic manner, and the origin of the periodic flips directly connects to the
unstable region. Interestingly, the toroidal field in the upper and lower halves always
have opposite signs, and the midplane horizontal field is very weak (and goes through
zero). We have also found that the overall mean field evolution can be almost exactly
reproduced from our quasi-1D simulation of paper I. An outflow is launched, whose mass
outflow rate is smaller than but the same order of magnitude to the rate from our run
R5b5H+ (see Table \ref{tab:stratruns}). Therefore, at a given time, the magnetic field
configuration can be considered physical for a magnetocentrifugal wind. However, since
the toroidal (hence radial) field constantly changes sign, the wind keeps oscillating
between radially inward and outward directions, a fact that is inconsistent with global
wind geometry, and reflects the limitation of the local shearing-box framework
\citep{BaiStone13a}. While the periodic field flips are likely physical phenomenon
inherent with the onset of the MRI, global simulations are necessary to determine the
fate of the outflow.

The onset of the MRI also leads to some level of turbulence, as seen from the bottom
panel of Figure \ref{fig:5AUprof}. Beyond the region where MRI operates, turbulent
motion largely results from passive response to the MRI activities, and the midplane
has the weakest level of turbulent motion. Despite different origins, the level of turbulence
is comparable to run R5b5H+, especially at the surface.

The fact that mean toroidal field periodically changes sign makes it ambiguous to
estimate the role of disk wind in transporting angular momentum (net wind-driven
accretion rate would be zero considering the periodic flips). Here we set it aside and
look at the radial transport of angular momentum from the Maxwell stress, as shown
in the top panel of Figure \ref{fig:5AUprof}. We see that Maxwell stress peaks
at about $|z|\sim4H$, but at a relatively low level. We estimate the total $\alpha$ to
be only about $4.5\times10^{-4}$, corresponding to accretion rate of
$\sim1.6\times10^{9}M_{\bigodot}$ yr$^{-1}$ using Equation (\ref{eq:accret}). This is
about an order of magnitude smaller than the expected level of
$10^{-8}M_{\bigodot}$ yr$^{-1}$.

We further performed run R5b4H$-$ with stronger net vertical field $\beta_0=10^4$.
Based on paper I, we expect the system to be stable to the MRI and develop a laminar
magnetocentrifugal wind. This is again confirmed using full 3D simulations, with the
general wind properties almost identical to the one obtained in paper I. In particular, our
full 3D run automatically obeys the reflection symmetry across the midplane, confirming
that solutions with enforced symmetry in paper I are generally physical. Note that
toroidal field is close to zero near the midplane as a result of the Hall term.
The level of random motion in our run R5B4H$-$ is systematically weaker than all other
runs, confirming its intrinsically laminar nature. One can read from Table \ref{tab:stratruns}
to obtain the Maxwell stress as well as the wind stress to derive the accretion rate
resulting from radial transport and wind, or directly look from Table 2 of paper I for more
accurate estimates. We see that radial transport is completely negligible compared with
wind-driven accretion rate, which gives the value of $\sim10^{-7}M_{\bigodot}$ yr$^{-1}$,
and is an order of magnitude more than sufficient.

In sum, it appears that for $B_{z0}<0$, while results from our shearing-box simulations
are likely robust, they also raise puzzling issues regarding the mechanism to transport angular
momentum. For relatively weak net vertical field ($\beta_0\sim10^5$), MRI sets in, leading
to a periodically oscillating outflow where based on shearing-box simulations we are
unable to tell if it drives angular momentum transport; but radial transport of angular
momentum by Maxwell stress appears too inefficient. For relatively strong net vertical field
($\beta_0\sim10^4$), the system unambiguously launches the magnetocentrifugal wind
which drives very rapid accretion with higher accretion rate than typically observed.
At this point it is unclear how the system can achieve accretion rate at the desired rate of
$\sim10^{-8}M_{\bigodot}$ yr$^{-1}$, an issue that can only be clarified from global
simulations.

\section[]{Simulations at Other Disk Radii}

In this section, we further perform simulations at two other locations, 15 AU and 60 AU,
from which we study the radial dependence of PPD gas dynamics and the role played by
the Hall effect. At each location, we perform four simulations with $\beta_0=10^4$ and
$10^5$ and different magnetic polarities, where all non-ideal MHD terms are turned on.

\subsection[]{Results from 15 AU}

\begin{figure}
    \centering
    \includegraphics[width=90mm]{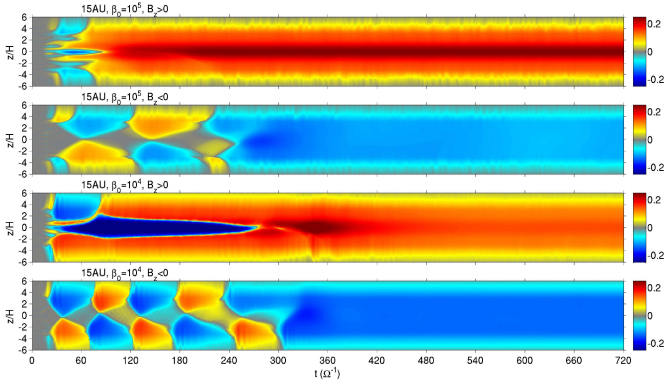}
  \caption{The time evolution for the vertical profile of horizontally averaged $B_y$ in
  our runs at 15 AU. Shown from top to bottom are runs R15b5H+, R15b5H$-$, R15b4+ and
  R15b4H$-$.}\label{fig:15AUhist}
\end{figure}

\begin{figure}
    \centering
    \includegraphics[width=70mm]{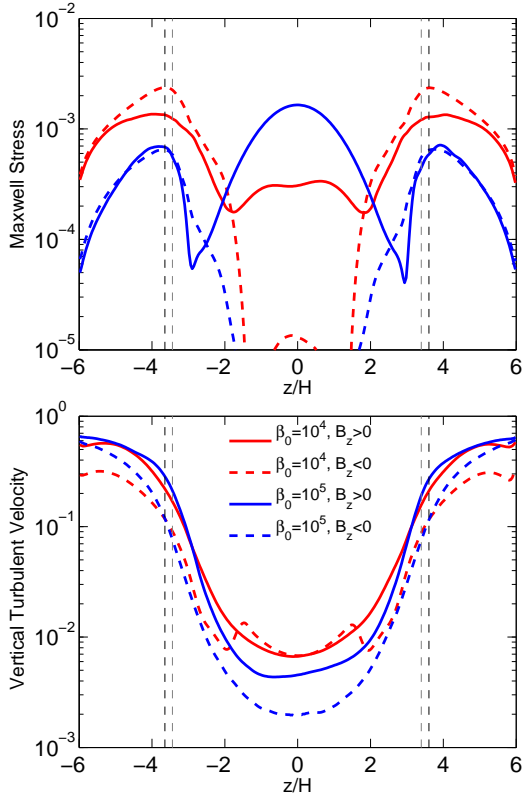}
  \caption{Vertical profiles of Maxwell stress (top) and vertical turbulent velocity
  (bottom) for all four runs at 15 AU, as marked in the legend. The vertical dashed
  lines mark the location where $Am=100$ in run R15b4H$-$ (dark) and R15b5H$-$
  (light).}\label{fig:15AUprof}
\end{figure}

At 15 AU, our quasi-1D simulations suggest laminar configuration for $B_{z0}>0$ with
$\beta_0=10^4$, but more turbulent situation is expected otherwise. In Figure
\ref{fig:15AUhist} we show the overall time evolution of the horizontally averaged toroidal
field. In Figure \ref{fig:15AUprof} we further show the time averaged profiles of Maxwell
stress and vertical turbulent velocity for all four runs, where the time averages are taken
from time $t=420\Omega^{-1}$ onward. We see that for all four runs, the system
eventually settle into a state where the large-scale toroidal field remains one sign across
the entire disk, hence the symmetry of the outflow would be undesirable for a global wind.
Nevertheless, we again set aside on the issue with symmetry and focus on other properties.

For $B_{z0}>0$ and comparing runs R15b5H$+$ with R15b4H$+$, it is counterintuitive to
notice from both Figures that stronger mean toroidal magnetic field is generated when the
net vertical field is weaker (R15b5H+), leading to stronger Maxwell stress around
disk midplane. Looking into the entire simulation data reveal that for run R15b4H+,
essential all the vertical magnetic flux is concentrated into a single thin shell, while the
rest of the radial zones have effective zero net vertical flux. As a result, magnetic field
amplification by the Hall shear instability is suppressed for the bulk of the disk. A strong
zonal flow is also formed with high density contrast of $30\%$ where shell of magnetic flux
locates at the density minimum.
The highly non-uniform distribution of magnetic flux also makes the gas dynamics in this
run deviate from the wind solution in paper I (see his Table 2).
On the other hand, for run R15b5H+, magnetic flux distribution is much more uniform,
leading to effective growth of horizontal magnetic field due to the Hall shear instability,
producing stronger Maxwell stress at disk midplane. Again, it is unclear at this point
how realistic the level of magnetic flux concentration is, hence the results shown here
should be treated with caution.

For $B_{z0}<0$, we see that the initial evolution of the mean toroidal field closely resembles
our run R5b5H$-$, with quasi-periodic flips and the top and bottom sides possesses opposite
sign of mean $B_y$. This is again because the MRI sets in in the layer where the Hall
Elsasser number transitions through order unity. Later on, field of one sign takes over and
dominates the entire disk. There are also MRI activities in the FUV layer, though the level
is weaker than their 30 AU counterpart (e.g., seen from the peak Maxwell stress). To some
extent, this location represents a transition between the 5AU and 30 AU cases, where in the
former MRI is triggered mainly in the Hall dominated layer while in the latter MRI is active
mainly in the FUV layer. As usual, the horizontal magnetic field is suppressed due to the Hall
effect, and most of the Maxwell stress originates from the FUV layer.

From the value of $\alpha^{\rm Max}$ and $T_{z\phi}^{\rm Max}$ listed in Table
\ref{tab:stratruns} and using Equation
(\ref{eq:accret}), we see that the net vertical magnetic flux has to be at least
$\beta_0=10^{4}$ in order for the accretion rate to reach levels comparable to
$10^{-8}M_{\bigodot}$ yr$^{-1}$. On the other hand, if magnetocentrifugal wind is
operating, the level of $T_{z\phi}^{\rm Max}$ from weak net vertical field with
$\beta_0=10^5$ is sufficient drive accretion rate above the desired level.
Overall, turbulent velocity is smallest at midplane either due to weak MRI turbulence
($B_{z0}>0$ with weak field) or induced random motion from MRI activities in the disk
surface ($B_{z0}<0$), similar to the 30 AU case.

\subsection[]{Results at 60 AU}

\begin{figure}
    \centering
    \includegraphics[width=90mm]{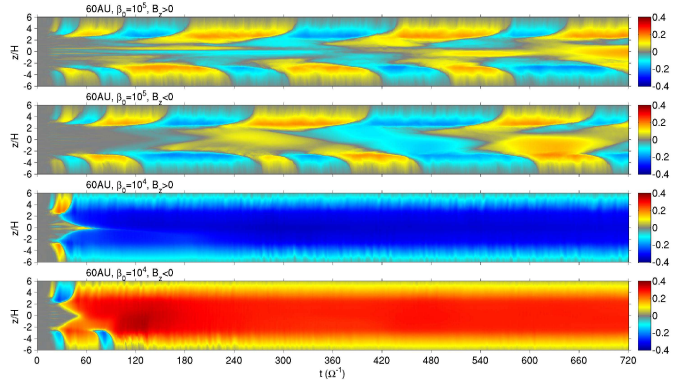}
  \caption{The time evolution for the vertical profile of horizontally averaged $B_y$ in
  our runs at 15 AU. Shown from top to bottom are runs R60b5H+, R60b5H$-$, R60b4+ and
  R60b4H$-$.}\label{fig:60AUhist}
\end{figure}

\begin{figure}
    \centering
    \includegraphics[width=70mm]{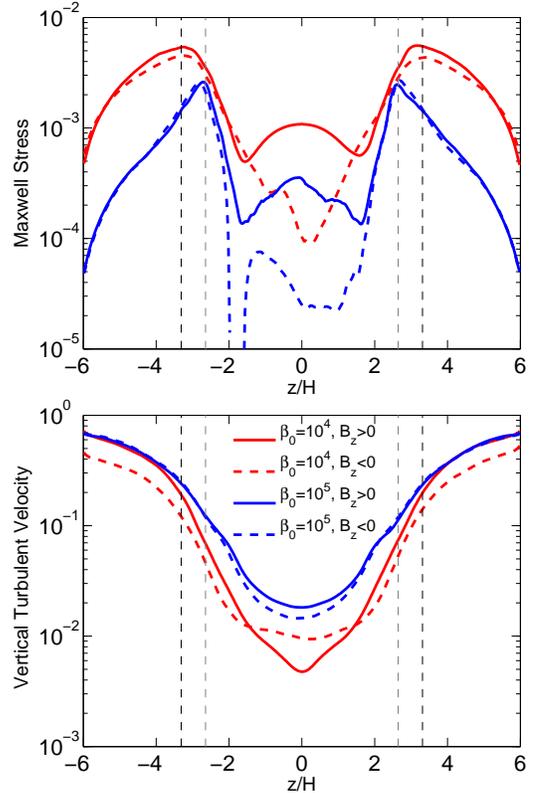}
  \caption{Vertical profiles of Maxwell stress (top) and vertical turbulent velocity
  (bottom) for all four runs at 60 AU, as marked in the legend. The vertical dashed
  lines mark the location where $Am=100$ in run R60b4H$-$ (dark) and R60b5H$-$
  (light).}\label{fig:60AUprof}
\end{figure}

At 60 AU, the relative importance of the Hall effect is reduced by a factor of $\sim2$ compared
with the 30 AU case (see Equation \ref{eq:Amchi}), and is only marginally important at disk
midplane. AD is the dominant effect in most regions of the disk. Also, given the approximately
constant penetration column density of the FUV ionization, it effectively penetrates deeper at
the more tenuous outer disk in terms of disk scale height. In Figure \ref{fig:60AUhist} we show
the overall time evolution of the horizontally averaged toroidal field. In Figure \ref{fig:60AUprof}
we further show the time averaged profiles of Maxwell stress and vertical turbulent velocity for
all four runs, where the time averages are taken from time $t=300\Omega^{-1}$ onward. The
general evolution of the system is in many ways similar to our focused study at 30 AU, where
MRI drives vigorous turbulence in the surface FUV layer, with the midplane region only weakly
turbulent. Here we mainly focus on the differences and the overall trend toward larger disk radii.

At $\beta_0=10^5$, dynamo activities constantly flip the mean toroidal field similar to but
appears more regular than the 30 AU case for both magnetic polarities. For $\beta_0=10^4$,
the dynamo is suppressed and the entire disk is dominated by a mean toroidal field with
a single sign. When $B_{z0}<0$, we do not observe the mean field changing sign as the
30 AU counterpart shown in Figure \ref{fig:30AUb4hist}. In fact the toroidal field in the entire
disk has the same sign throughout the saturated state of the simulation hence we do not
expect this sign flip to occur. We speculate that the flip we observed at 30 AU is associated
with the relatively strong Hall effect at the disk midplane, but it is unlikely to occur toward the
outer disk as the Hall effect becomes less dominant.

At 60 AU, the contrast in Maxwell stress between the $B_{z0}>0$ and $B_{z0}<0$ cases at
disk midplane is still very evident. Level of turbulence is found to be higher for runs with
weaker net vertical field $\beta_0=10^5$, which may be due to the fact that in runs with
$\beta_0=10^4$, turbulent motion is limited by the relatively strong large-scale toroidal
field, but it may also be due to strong concentration of magnetic flux into thin shells where
a large fraction of the simulation domain has effectively zero net vertical flux. 

Deeper penetration of FUV ionization allows the MRI to be fully active over thicker surface
layers, hence the Maxwell stress profiles at disk surface at fixed $\beta_0$ is higher than
the their 30AU counterparts, giving larger values of $\alpha^{\rm Max}$. Again, we find that
for the Maxwell stress alone to drive accretion rate of $\sim10^{-8}M_{\bigodot}$ yr$^{-1}$, the
net vertical flux needs to be $\beta_0\sim10^4$ or stronger. The magnetocentrifugal wind,
if operating in the outer disk, would drive accretion with rate
$\sim0.4-4\times10^{-8}M_{\bigodot}$ yr$^{-1}$ for $\beta_0=10^{5}$ to $10^4$.

\section[]{Summary and Discussions}\label{sec:summary}

\subsection{Summary}

In this work, we have studied the gas dynamics of PPDs focusing on regions toward the
outer disk (from 5-60 AU), taking into account all non-ideal MHD effects in a self-consistent
manner. In these regions, the Hall effect generally dominates near the disk midplane, and
ambipolar diffusion (AD) plays an important role over a more extended region across disk
height, and the very surface layer behaves in the ideal MHD regime due to FUV ionization.
In the presence of the Hall effect, the gas dynamics depends on the polarity of the large-scale
vertical/poloidal magnetic field ($B_{z0}$) threading the disk relative to the rotation axis
(along $\hat{z}$). Since the relative importance of the Hall effect to AD gets progressively
weaker with increasing disk radius, we estimate based on the MMSN disk model
that the Hall-effect controlled polarity dependence extends to about 60 AU.

We first conducted unstratified MRI simulations including both the Hall effect and AD. We
find that at conditions expected in the outer region of PPDs (midplane plasma $\beta_0$
for the net vertical field being $10^{4-5}$), MRI leads to turbulence when $B_{z0}>0$ but can
not be self-sustained for $B_{z0}<0$. For $B_{z0}>0$, the level of MRI turbulence is of the
order $\alpha\sim10^{-3}$ (with AD Elsasser number $Am=1$). We confirm that strong
zonal field configuration of \citet{KunzLesur13} can be achieved with sufficiently strong Hall
effect, and find that in the mean time it leads to strong zonal flows. In addition, numerical
resolution of $24$ cells per $H=c_s/\Omega$ is in general sufficient to resolve the bulk
properties of the MRI turbulence.

We then focused on self-consistent stratified MRI simulations at fixed disk radius, with main
results summarized as follows.

At relatively small disk radius ($\sim5$ AU), and for $B_{z0}>0$, we confirm and justify
the results from paper I that the system launches a strong magnetocentrifugal wind, and is
able to achieve a physical wind geometry, with the horizontal magnetic field flips exactly at
disk midplane. While Maxwell stress is enhanced due to the Hall shear instability, accretion
is largely driven by the wind and proceeds primarily through the midplane. In addition, the
midplane region is weakly turbulent which is likely resulting from the flip of relatively strong
horizontal magnetic field. The turbulent motion gets amplified toward disk surface as gas
density drops.

For $B_{z0}<0$, our full 3D simulations confirm results from paper I that the system is
unstable to the MRI in thin Hall-dominated layers when net vertical field is relatively weak
($\beta_0=10^5$). This results in periodic flips of large-scale horizontal magnetic field over
time with a radially oscillating disk outflow/wind whose fate and whether it drives accretion
are uncertain based on shearing-box simulations. Radial transport of angular momentum
by Maxwell stress is found to be too inefficient by an order of magnitude. A stable
magnetocentrifugal wind with physical wind geometry can be achieved with stronger net
vertical field ($\beta_0=10^4$), which very efficiently drives accretion with
$\dot{M}\gtrsim10^{-7}M_{\bigodot}$ yr$^{-1}$. It is uncertain whether and how the system
can achieve the typically observed rate of $10^{-8}M_{\bigodot}$ yr$^{-1}$.

At relatively large disk radius ($\sim30$ AU), we find that the Hall effect mainly affects the
Maxwell stress at disk midplane, with $B_{z0}>0$ ($B_{z0}<0$) giving enhanced (reduced)
stress similar to those found at the inner disk (paper I, \citealp{Lesur_etal14}). Nevertheless,
strongest Maxwell stress results from vigorous MRI turbulence in the surface layer due to
FUV ionization \citep{PerezBeckerChiang11b,Simon_etal13b}. While self-sustained MRI is
expected at disk midplane when $B_{z0}>0$ but not when $B_{z0}<0$, the level of
turbulence in these cases appears very similar, with vertical turbulent velocity of the order
$\delta v_z\sim0.01-0.03c_s$. The turbulent motion in the latter case is largely induced from
stronger turbulence in the surface layer analogous to the conventional ``Ohmic dead zone"
picture (e.g., \citealp{FlemingStone03}). Overall, the gas dynamics in the outer regions of
PPDs show clear layered structure consisting of highly turbulent surface FUV ionization layer
with strong Maxwell stress and weakly turbulent midplane region due to a combination of AD,
the Hall effect and large-scale magnetic field structure.

We find that for relatively weak field ($\beta_0=10^5$), MRI dynamo leads to repeated flips
of large-scale toroidal field, with very irregular cycles. Dynamo activities tends to be
suppressed for stronger fields ($\beta_0=10^4$). Our simulations also show secular
behavior on the evolution of mean toroidal field, especially in simulations at $30$ AU. This
is to a certain extent related to the limitations of shearing-box, since the net vertical magnetic
flux ought to connected to infinity but gets truncated by the vertical boundary condition
without reaching all the critical points (e.g., \citealp{Fromang_etal13}).

We also find that most of our simulations show strong concentration of vertical magnetic flux
into a thin azimuthal shell at certain radial location, while the rest of the regions have close to
zero net vertical flux. The concentration is generally stronger in simulations with stronger net
vertical field ($\beta_0=10^4$) and toward outer disk radii ($\gtrsim15$ AU). The
concentration differs from the zonal field due to the Hall effect \citep{KunzLesur13}, but
appears to be generic in shearing-box simulations with net vertical magnetic flux and turbulence.
Accompanied with magnetic flux concentration is enhanced density variation across the radial
domain, with most flux is concentrated in low density regions. While this is likely the origin of
enhanced zonal flow from shearing-box simulations \citep{SimonArmitage14}, it remains to
clarify the physics of magnetic flux concentration, and study its saturation amplitude in global
context.

While all our simulations launch disk outflows, it is uncertain whether such outflows (at
$\gtrsim15$ AU) can be incorporated into a global magnetocentrifugal wind due to
MRI dynamo and symmetry issues \citep{BaiStone13a}, but if they do, the level of net vertical
flux $\beta_0=10^{5}$ and stronger are generally sufficient to drive accretion at desired level
of $10^{-8}M_{\bigodot}$ yr$^{-1}$. On the other hand, to rely on purely radial transport of
angular momentum by Maxwell and Reynolds stresses, the level of net vertical field must be
$\beta_0=10^4$ or stronger assuming MMSN disk model. This level of field translates to
physical field strength according to
\begin{equation}
B=18.6\beta^{-1/2}R_{\rm AU}^{-13/8}\ {\rm G}\ .\label{eq:Bfield}
\end{equation}
For reference, we find for $\beta_0=10^4$, $B_{z0}\sim0.7$ mG at 30 AU.

\subsection{Discussions}

Combining the results from this paper and paper I together, we see that the Hall effect
has major influence to the disk dynamics toward inner region of PPDs ($\lesssim15$
AU) where polarity dependence  is most prominent in determining the wind properties,
stability to the MRI, and the amplification/reduction of horizontal magnetic field in the Hall
dominated regions. The Hall effect also affect the stability to the MRI in the midplane
region of the outer disk though it is not quite significant in setting the level of turbulent
motions. Overall, it is likely that wind-driven accretion dominates the inner disk while
accretion can be largely driven by the MRI at surface FUV layer in the outer disk, as
outlined in the discussion of \citet{Bai13}, which incorporated numerical simulation
results without the Hall effect \citep{BaiStone13b,Simon_etal13b}.  On the other hand,
detailed behavior in the inner disk region, as well as the transition from the largely
laminar inner region to the MRI turbulent outer disk region, are expected to have strong
polarity dependence due to the Hall effect, as summarized in the previous subsection,
and also in paper I and \citet{Lesur_etal14}.

Several observational consequences are expected based on our current simulation
results. First, the fact that the inner disk launches a magnetocentrifugal wind can be
detectable through gas tracers. In fact, signatures of low velocity disk outflow have
been routinely inferred from blue-shifted emission line profiles such as from CO, OI
and NeII lines (e.g., \citealp{Hartigan_etal95,Pascucci_etal09,Pontoppidan_etal11,
Herczeg_etal11,Sacco_etal12,Rigliaco_etal13}). While conventionally interpreted as
signatures of photo-evaporation (e.g., \citealp{Gorti_etal09,Owen_etal10}),
magnetocentrifugal wind is likely to produce similar signatures, since they possess low
velocities near the launching point before getting strongly accelerated and diluted. In
reality, both mechanisms are likely to contribute to launching the outflow due to the
combination of UV radiative transfer and photochemistry, thermodynamics, and magnetic
fields. We note that a pure photoevaporative wind is likely to be {\it angular-momentum}
conserving since the radial driving force does not exert any torque to the outflow, while a
magnetocentrifugal wind is more likely to be {\it angular-velocity} conserving near the
base of the wind where the gas is forced to move along supra-thermal magnetic fields
anchored to the disk (e.g., \citealp{Spruit96}). Searching for distinguishable signatures
between the two scenarios would be important for understanding the nature of the
observed disk outflows.

Second, we expect the level of turbulence in the outer disk to be layered, where the
level of turbulence is expected to be of the order $\delta v_z\sim10^{-2}c_s$ at midplane
and increases to near sonic level toward disk surface (the full turbulent velocity is further
higher). Empirical constraint on the level of turbulence in the outer region of PPDs has
already been reported based on the turbulent line width of the CO (3-2) transition
\citep{Hughes_etal11}. This line is optically thick and probes the disk surface layer with
line width constrained to be $\lesssim10-40\%$ of sound speed, consistent with a fully
turbulent surface layer. With superb sensitivity and resolution, ALMA is expected to
constrain the variations of turbulence level at different disk heights using different line
tracers, which will provide direct evidence of layered structure of the outer PPDs.

Third, the weakly turbulent outer disk with toroidal dominated field configuration may
lead to grain alignment and dust polarization \citep{ChoLazarian07}. We have found that
in the outer disk ($\gtrsim30$ AU), the net vertical field needs to be $\beta_0\sim10^4$ or
stronger for Maxwell stress to drive accretion rate of $10^{-8}M_{\bigodot}$ yr$^{-1}$. For
such level of net vertical field, we see that the MRI dynamo is suppressed, and the entire
field is dominated by a large-scale toroidal magnetic field, whose strength at disk midplane
corresponds to plasma $\beta\sim10-20$ (e.g., see Figure \ref{fig:60AUhist}). Using Equation
(\ref{eq:Bfield}), we find the midplane toroidal field can be at least $\sim3-8$ mG at
$60-100$ AU. Based on Equation (1) of \citet{Hughes_etal09}, and using the MMSN disk
model at midplane with grain size of $10-100\mu$m and dust aspect ratio $s=3$, we find
the critical strength for grain alignment to occur is $\sim1-40$ mG at 60-100 AU. While
there are large theoretical uncertainties, we see that the match is marginal, and the field
strength in the outer disk can either be just enough for promoting grain alignment, or
a little too weak to align the grains. Several observational attempts to search for dust
polarization in Class II disks have failed \citep{Hughes_etal09,Hughes_etal13}. Very
recently, however, successful detection of dust polarization toward younger sources have
been reported, with inferred field configuration resembling large scale toroidal field
(\citealp{Rao_etal14}, Stephens et al., in preparation). This might indicate that disk
magnetic field fades over time. Again, future dust polarization observations by ALMA will
likely provide better constraints on the geometry, strength and evolution of disk magnetic
fields.

From this work together with paper I, we have explored the main parameter space on
the gas dynamics of PPDs using local shearing-box simulations. There are other
unexplored parameters and uncertainties including the abundance and size distribution
of grains, where tiny grains such as polycyclic-aromatic-hydrocarbons may reduce the
importance of the Hall effect and AD hence promote the MRI \citep{Bai11b}. Also, the
cosmic-ray ionization rate may be reduced and modulated by stellar wind/disk wind
\citep{Cleeves_etal13}, the X-ray luminosity can be highly variable due to stellar flares
\citep{Wolk_etal05,IlgnerNelson06c}, and FUV photons may be shielded by the dust in
the disk wind from the inner disk \citep{BansKonigl12}. It is likely that grain abundance
and FUV ionization are more sensitive parameters (\citealp{BaiStone13b,Simon_etal13b},
paper I), and X-ray ionization is less sensitive but also important (\citealp{Bai11a}, paper I).

Probably the largest uncertainties in our work come from the use of local shearing-box
framework, and there are several outstanding issues resulting from the net vertical magnetic
flux. With net vertical flux, it has been well known that the properties of the disk outflow is
not well characterized in shearing-box simulations largely because the vertical gravitational
potential is ever-increasing in the local approximation \citep{Fromang_etal13,BaiStone13b}.
The issues related to the symmetry and fate of the outflow is notorious
\citep{BaiStone13a,BaiStone13b}. Moreover, the evolution of large-scale magnetic field
can be affected by the vertical outflow boundary condition. In this paper, we further identify
the issue with the concentration of vertical magnetic flux into thin shells which resides in
low-density regions in the zonal flow.
Global disk simulations with vertical stratification and net vertical
magnetic flux have recently been carried out \citep{SuzukiInutsuka14}, yet many of these
issues remain not quite addressed due to limited domain size in the $\theta$ dimension.
In the future, it is crucial to perform global simulations with sufficiently large vertical
domain to accommodate the disk outflow/wind, and fine resolution in the disk to resolve
the disk microphysics. In this way, these critical issues can potentially and ultimately be
appropriately addressed.

\acknowledgments

I thank Jim Stone for helpful discussions and useful comments to the draft.
This work is supported from program number HST-HF-51301.01-A provided
by NASA through a Hubble Fellowship grant from the Space Telescope Science
Institute awarded to XN.B, which is operated by the Association of Universities for
Research in Astronomy, Incorporated, under NASA contract NAS5-26555.
Computation for this work was performed on Stampede at Texas Advanced Computing
Center through XSEDE grant TG-AST140001, and on Kraken at National Institute for
Computational Sciences through XSEDE grant TG-AST130048.

\bibliographystyle{apj}

\label{lastpage}
\end{document}